\documentclass[
aps,
prd,
twocolumn,
showpacs,
superscriptaddress,
nofootinbib,
floatfix
]{revtex4-2}

\usepackage{amsmath,amssymb,amsfonts}
\usepackage{bm}
\usepackage{slashed}

\usepackage{graphicx}
\usepackage{booktabs}
\usepackage{dcolumn}
\usepackage{subcaption}

\usepackage{caption}
\usepackage{ragged2e}

\usepackage[
colorlinks=true,
linkcolor=blue,
citecolor=blue,
urlcolor=blue
]{hyperref}

\usepackage{xcolor}

\begin{document}

%%%%%%%%%%%%%%%%%%%%%%%%%%%%%%%%%%%%%%%%%%%%%%%%%%%%%
\title{Open-Charm Vector Mesons in Hot and Dense  Matter}

\author{N.~Er}
\thanks{Corresponding author}
\email{nuray@ibu.edu.tr}
\affiliation{Department of Physics, Bolu Abant \.{I}zzet Baysal University,
  G\"{o}lk\"{o}y Campus, 14030 Bolu,  T\"{u}rkiye}

\author{K.~Azizi}
\thanks{Corresponding author}
\email{kazizi@dogus.edu.tr}
\affiliation{Department of Physics, Faculty of Engineering and Natural Sciences, Dogus University,
  Dudullu-\"{U}mraniye, 34775  Istanbul, T\"{u}rkiye}

\date{\today}

%%%%%%%%%%%%%%%%%%%%%%%%%%%%%%%%%%%%%%%%%%%%%%%%%%%%%
\begin{abstract}
%%%%%%%%%%%%%%%%%%%%%%%%%%%%%%%%%%%%%%%%%%%%%%%%%%%%%

We investigate the in-medium properties of the open-charm vector mesons
$D_s^{*}$ and $D^{*}$ in hot and dense nuclear matter within the
framework of finite-temperature and finite-density QCD sum rules. The
analysis incorporates temperature- and density-dependent quark and
gluon condensates together with an in-medium continuum threshold
constrained by the light-quark condensate. By solving the resulting QCD
sum rules, we determine the in-medium masses and leptonic decay
constants of the $D_s^{*\pm}$ and $D^{*\pm}$ mesons over a broad region
of the $(T,\rho)$ plane. Both vector mesons undergo substantial
in-medium softening, with their masses and leptonic decay constants
decreasing as the baryon density increases. The masses exhibit a
non-monotonic dependence on baryon density, whereas the leptonic decay
constants decrease monotonically throughout the investigated density
range. Increasing temperature generally weakens the density-induced
modifications, although baryon density remains the dominant driver of
the in-medium evolution. The largest mass shifts occur at
intermediate-to-high densities, reaching approximately
$-413~\mathrm{MeV}$ for the $D_s^{*-}$ meson and
$-207~\mathrm{MeV}$ for the $D^{*-}$ meson, while the leptonic decay
constants are reduced by more than $68\%$ in both channels at the
highest densities considered. We further investigate the
particle--antiparticle splittings of the masses and leptonic decay
constants induced by finite baryon density. Finite baryon density lifts
the vacuum degeneracy between the charge-conjugate states, while
increasing temperature generally suppresses the resulting asymmetries.
Although the strange and non-strange channels exhibit similar
qualitative behavior, quantitative differences emerge in both the
in-medium modifications and the particle--antiparticle splittings. These
results provide quantitative predictions for open-charm vector mesons in
hot and dense QCD matter and may serve as theoretical benchmarks for
future heavy-flavor measurements at FAIR-CBM, NICA, and J-PARC.

\end{abstract}

\maketitle

%%%%%%%%%%%%%%%%%%%%%%%%%%%%%%%%%%%%%%%%%%%%%%%%%%%%%
\section{Introduction}
%%%%%%%%%%%%%%%%%%%%%%%%%%%%%%%%%%%%%%%%%%%%%%%%%%%%%
The behavior of strongly interacting matter under extreme thermodynamic
conditions remains one of the central problems of Quantum Chromodynamics
(QCD). At sufficiently high temperatures or baryon densities, QCD predicts
profound modifications of the hadronic vacuum, culminating in a
transition to the deconfined quark--gluon plasma (QGP), where hadronic
bound states dissolve and quarks and gluons become deconfined
degrees of freedom~\cite{Shuryak,Fukushima2010ThePD,Karsch2002}. Understanding this transition and the
associated evolution of hadronic properties is essential for describing
the early Universe, the interior of compact stars, and the strongly interacting medium produced in relativistic heavy-ion collisions. It also offers a valuable framework for exploring the nonperturbative
structure of QCD, particularly the mechanisms of color confinement and
the in-medium evolution of chiral symmetry~\cite{HatsudaLee1992,
RappWambach2000,Hayashigaki:2000es}.

Modern accelerator facilities such as the Relativistic Heavy Ion
Collider (RHIC) and the Large Hadron Collider (LHC) recreate strongly
interacting matter at extremely high temperatures and relatively low
baryon densities, providing experimental access to the corresponding
region of the QCD phase diagram. These conditions resemble those
prevailing during the earliest stages of the Universe following the Big
Bang~\cite{Arsene2005,Adams2005,ALICE:2010khr}. The region of the QCD phase diagram characterized by large baryon
densities and moderate temperatures remains considerably less explored
than its high-temperature counterpart. Understanding the properties of
matter under such conditions is essential for describing the dense
interiors of neutron stars and represents one of the major goals of the
next generation of heavy-ion facilities. Experimental programs at the
Facility for Antiproton and Ion Research (FAIR/GSI), the Japan Proton
Accelerator Research Complex (J-PARC), and the Nuclotron-based Ion
Collider fAcility (NICA) are expected to provide unprecedented
information on strongly interacting matter in this previously
inaccessible region of the QCD phase
diagram~\cite{Ablyazimov:2017guv,KEKELIDZE2016846,Sako:2014xph,
Baym:2017whm,Bzdak2020,Friman2011}.

Within this context, heavy mesons containing a single heavy quark
constitute particularly sensitive probes of the surrounding strongly
interacting medium~\cite{Hayashigaki:2000es,Hilger2011}. Their in-medium
modifications influence heavy-quark transport, the production and
suppression of heavy quarkonia, and the possible formation of exotic
hadronic states in nuclear matter. Among these systems, open-charm
mesons are especially attractive because they directly probe the
interaction of charm quarks with the surrounding medium and are
accessible in ongoing and future experimental programs at FAIR, NICA,
J-PARC, RHIC, and the LHC~\cite{Tolos2009}.

Unlike the pseudoscalar $D_s$ and $D$ mesons, the corresponding vector
states are interpolated by the vector current and receive different
perturbative and nonperturbative contributions within the operator
product expansion (OPE). Their in-medium properties therefore cannot be
inferred directly from the pseudoscalar sector and must be investigated
independently. Among the open-charm vector mesons, the $D_s^*$ and
$D^*$ states provide an excellent opportunity to explore the influence
of strange- and light-quark dynamics in heavy--light systems. Since the
strange-quark condensate exhibits a different in-medium evolution from
its light-quark counterpart, a comparative investigation of the
$D_s^*$ and $D^*$ channels offers direct insight into
flavor-dependent medium effects. Moreover, the masses and leptonic decay
constants of these mesons constitute important inputs for
phenomenological studies of heavy-ion collisions, open-charm transport,
and the in-medium behavior of charmed hadrons.

A variety of theoretical approaches have been developed to investigate
the in-medium properties of heavy mesons, including effective hadronic
models, chiral effective theories, lattice QCD simulations, and QCD sum
rules~\cite{Klingl1999,Hilger2009,Morita2008}. Among these methods, QCD
sum rules provide a well-established nonperturbative framework that
relates hadronic observables directly to the underlying QCD condensates
by combining the operator product expansion with quark--hadron
duality~\cite{Furnstahl:1992ux,Shifman:1978bx,Colangelo:2000dp,Cohen:1991nk}. Within this framework, changes in hadronic observables can be traced back to the medium dependence of the underlying QCD condensates,
thereby establishing a direct connection between measurable hadronic
properties and the nonperturbative structure of QCD at finite
temperature and baryon density, including the gradual restoration of
chiral symmetry~\cite{HatsudaLee1992,RappWambach2000}. Owing to this
feature, QCD sum rules have become one of the principal nonperturbative
tools for investigating the in-medium behavior of both light- and
heavy-flavor hadrons~\cite{HatsudaLee1992,Hayashigaki:2000es,Hilger2009}.

Under hot and dense conditions, Lorentz invariance is
explicitly broken by the presence of the medium four-velocity
$u^\mu$, giving rise to additional operators in the operator product
expansion and to new medium-specific condensates beyond their vacuum
counterparts~\cite{Bochkarev1986,HatsudaLee1992,Ayala2017}. As a result, hadronic observables acquire a nontrivial dependence on temperature and baryon density, reflecting the medium evolution of the
underlying QCD condensates.

Despite the considerable theoretical effort devoted to the in-medium
behavior of heavy mesons using QCD sum rules, effective hadronic
approaches, and quark-based models~\cite{Hayashigaki:2000es,
Hilger2011,Azizi2014,Cobos-Martinez:2025iqg}, the simultaneous impact
of finite temperature and baryon density on the open-charm vector
mesons $D_s^*$ and $D^*$ has not yet been systematically established. Owing to their different light-quark flavor
content, a comparative investigation of these two systems provides an
opportunity to clarify the role of strangeness in the in-medium
modification of heavy--light vector mesons. Furthermore, besides the
medium dependence of the masses and leptonic decay constants, the
particle--antiparticle mass and decay-constant splittings generated by
finite baryon density constitute additional observables for probing the
breaking of charge-conjugation symmetry in nuclear matter.

In this work, we present a finite-temperature and finite-density QCD
sum-rule analysis of the open-charm vector mesons $D_s^*$ and $D^*$ in
nuclear matter.The analysis incorporates medium-dependent quark, gluon, and mixed
condensates within the operator product expansion to determine the
in-medium masses and leptonic decay constants. Particular emphasis is placed on
elucidating how finite temperature and baryon density modify the
masses, leptonic decay constants, and particle--antiparticle splittings
of both open-charm vector-meson channels within a common QCD sum-rule
framework. The results presented here provide quantitative benchmarks for future
theoretical studies and may assist in the interpretation of
open-charm observables measured at FAIR, NICA, and J-PARC.

%%%%%%%%%%%%%%%%%%%%%%%%%%%%%%%%%%%%%%%%%%%%%%%%%%%%%
\section{In-Medium QCD Duality Framework under Extreme Conditions}
\label{sec:Method}
%%%%%%%%%%%%%%%%%%%%%%%%%%%%%%%%%%%%%%%%%%%%%%%%%%%%%

The in-medium QCD sum-rule approach provides a well-established
nonperturbative framework for relating hadronic observables to the
underlying degrees of freedom of QCD through quark--hadron duality.
Its central ingredient is the two-point correlation function, which is
evaluated independently in terms of hadronic degrees of freedom and
within the OPE. Under finite-temperature
and finite-density conditions, the vacuum expectation values of local
operators are replaced by their corresponding in-medium expectation
values, thereby incorporating thermal and density effects directly into
the correlation function. Matching the hadronic and QCD
representations allows the in-medium masses and leptonic decay
constants of the vector mesons to be determined. In this section, we outline the theoretical ingredients of the
finite-temperature and finite-density QCD sum-rule formalism employed
throughout this work. We first construct the hadronic and QCD sides of
the two-point correlation function, then derive the corresponding sum
rules, and finally specify the medium-dependent condensates and input
parameters used in the numerical analysis.

%%%%%%%%%%%%%%%%%%%%%%%%%%%%%%%%%%%%%%%%%%%%%%%%%%%%%
\subsection{In-Medium Hadronic Representation}
%%%%%%%%%%%%%%%%%%%%%%%%%%%%%%%%%%%%%%%%%%%%%%%%%%%%%

The analysis starts from the in-medium two-point correlation function
evaluated in hot and dense nuclear matter. Within the QCD sum-rule
framework, this correlator provides the connection between the hadronic
degrees of freedom and the underlying quark--gluon dynamics, thereby
allowing the medium modifications of the vector-meson masses and
leptonic decay constants to be determined. The in-medium two-point correlation function is given by
\begin{equation}\label{eq:CF}
\Pi_{\mu\nu}(T,\rho)
= i\!\int\!d^4x\,e^{ip\cdot x}
\langle\mathbf{T}\{J_\mu(x)J_\nu^\dagger(0)\}\rangle_{T,\rho},
\end{equation}
where $\mathbf{T}$ is the time-ordering operator and
$J_\mu(x)$ is the interpolating current associated with the quantum
numbers of the vector meson, and
$\langle\cdots\rangle_{T,\rho}$ represents the thermal average over the
hot and dense nuclear medium.

For a heavy-light vector meson with valence quark content
$\bar{\psi}_1\psi_2$, the interpolating current is chosen as
\begin{equation}\label{current}
J_\mu(x)=\bar{\psi}_1(x)\gamma_\mu\psi_2(x).
\end{equation}
In the present analysis, we consider the open-charm vector mesons
$D^{*+}(c\bar d)$,
$D^{*-}(\bar c d)$,
$D_s^{*+}(c\bar s)$,
and
$D_s^{*-}(\bar c s)$,
following the Particle Data Group (PDG)
classification~\cite{PDG:2024}. The corresponding interpolating
currents are obtained by assigning the appropriate quark fields to
$\psi_1(x)$ and $\psi_2(x)$ in Eq.~(\ref{current}).

The hadronic representation is obtained by inserting a complete set of
physical intermediate states into the correlation function and isolating the
ground-state contribution. This leads to
\begin{equation}\label{eq:HAD1}
\Pi_{\mu\nu}^{\rm HAD}(T,\rho)
=
\frac{
\langle J_\mu|D_{(s)}^*(T,\rho)\rangle
\langle D_{(s)}^*(T,\rho)|J_\nu^\dagger\rangle
}
{
m_{D_{(s)}^*}^{*2}(T,\rho)-p^{*2}
}
+\cdots,
\end{equation}
where
$m_{D_{(s)}^*}^{*}(T,\rho)$ is the in-medium pole mass of the
$D_{(s)}^*$ vector meson.
The effective four-momentum is defined as
$p_\mu^{*}=p_\mu-\Sigma_{\mu,v}$,
where
$\Sigma_{\mu,v}
=\Sigma_vu_\mu+\Sigma_v'p_\mu$.
Since the contribution of $\Sigma_v'$ is numerically small, it is
neglected throughout this work. Here,
$\Sigma_v$ denotes the vector self-energy and
$u_\mu$ is the four-velocity of the medium.
The calculations are performed in the rest frame of nuclear matter,
where
$u_\mu=(1,\mathbf{0})$.
The omitted terms account for contributions from higher resonances and the continuum.

The leptonic decay constant constitutes an important hadronic parameter
that characterizes the overlap between the interpolating current and the
physical meson state. In the framework of in-medium QCD sum rules, its
medium dependence provides complementary information to the mass shift,
reflecting modifications of the internal quark--gluon structure of the
meson induced by the surrounding nuclear environment. The in-medium leptonic
decay constant, $f_{D_{(s)}^*}^{*}(T,\rho)$, is defined through the
matrix element
\begin{equation}\label{DCons}
\langle J_\mu|D_{(s)}^*(T,\rho)\rangle
=
f_{D_{(s)}^*}^{*}(T,\rho)\,
m_{D_{(s)}^*}^{*}(T,\rho)\,
\epsilon_\mu,
\end{equation}
where $\epsilon_\mu$ denotes the polarization vector of the vector meson.

Substituting Eq.~(\ref{DCons}) into Eq.~(\ref{eq:HAD1}) yields
\begin{eqnarray}\label{eq:HAD2}
\Pi_{\mu\nu}^{\rm HAD}(T,\rho)
&=&
-\frac{
f_{D_{(s)}^*}^{*}(T,\rho)
m_{D_{(s)}^*}^{*2}(T,\rho)
}
{
m_{D_{(s)}^*}^{*2}(T,\rho)-p^{*2}
}
\nonumber\\
&&\times
\left(
-g_{\mu\nu}
+
\frac{p_\mu^{*}p_\nu^{*}}
{m_{D_{(s)}^*}^{*2}(T,\rho)}
\right) \nonumber\\
&&
+\cdots .
\end{eqnarray}

Using the explicit expression for the effective momentum
$p_\mu^{*}$,
the hadronic representation can be rewritten as
\begin{eqnarray}\label{eq:HAD3}
\Pi_{\mu\nu}^{\rm HAD}(T,\rho)
&=&
-\frac{
f_{D_{(s)}^*}^{*}(T,\rho)
}
{
p^2-\kappa^2
}
\Big[
-g_{\mu\nu}
m_{D_{(s)}^*}^{*2}(T,\rho)
\nonumber\\
&&
+ \, p_\mu p_\nu
-\Sigma_v
(p_\mu u_\nu+p_\nu u_\mu)
+\Sigma_v^2u_\mu u_\nu
\Big] \nonumber\\
&&
+\cdots,
\end{eqnarray}
where
\[
\kappa^2
=
m_{D_{(s)}^*}^{*2}(T,\rho)
-\Sigma_v^2
+2p_0\Sigma_v,
\]
with
$p_0=p\!\cdot\!u$
denoting the quasi-particle energy.

Besides the ground-state pole, the hadronic representation contains
contributions from excited resonances and the continuum, whose detailed
spectral structure is not explicitly known.
To reduce the contributions from higher resonances and the continuum
while improving the convergence of the operator product expansion on
the QCD side, a Borel transformation is applied,
\begin{eqnarray}\label{eq:Borel}
\hat{\mathcal{B}}_{p^2}
\equiv
\lim_{\substack{-p^2,n\rightarrow\infty\\-p^2/n=M^2}}
\frac{(-p^2)^{n+1}}{n!}
\left(\frac{d}{dp^2}\right)^n,
\end{eqnarray}
where $M^2$ is the Borel mass parameter.
After the transformation, the hadronic side takes the form
\begin{eqnarray}\label{eq:HAD4}
\Pi_{\mu\nu}^{\rm HAD}(T,\rho)
&=&
f_{D_{(s)}^*}^{*}(T,\rho)
e^{-\kappa^2/M^2}
\Big[
-g_{\mu\nu}
m_{D_{(s)}^*}^{*2}(T,\rho)
\nonumber\\
&&
+p_\mu p_\nu
-\Sigma_v
(p_\mu u_\nu+p_\nu u_\mu)
+\Sigma_v^2u_\mu u_\nu
\Big] \nonumber\\
&&
+\cdots .
\end{eqnarray}

%\FloatBarrier
%%%%%%%%%%%%%%%%%%%%%%%%%%%%%%%%%%%%%%%%%%%%%%%%%%%%%
\subsection{In-medium QCD Representation and the Coupled Sum Rules}
%%%%%%%%%%%%%%%%%%%%%%%%%%%%%%%%%%%%%%%%%%%%%%%%%%%%%

The QCD representation of the correlation function is constructed
within the OPE. Substituting the
interpolating current given in Eq.~\eqref{current} into
Eq.~\eqref{eq:CF} and contracting the quark fields according to
Wick's theorem, one obtains
\begin{equation}\label{eq:QCD1}
\Pi_{\mu\nu}^{\rm QCD}(T,\rho)
=
i\int d^4x\,e^{ip\cdot x}
\,
\mathrm{Tr}
\!\left[
S_{\psi_1}^{ab}(x)\gamma_\nu
S_{\psi_2}^{ba}(-x)\gamma_\mu
\right],
\end{equation}
where
$S_{\psi_{1,2}}^{ab}(x)$
denote the full in-medium quark propagators and
$a,b$
are color indices.

Within the fixed-point gauge and in the chiral limit, the coordinate-space
representation of the light-quark propagator at finite temperature and
baryon density is given by
\begin{align}\label{eq:lightprop}
S_q^{ab}(x)
  &= \left(1 - n_F\right)
     \left[
       \frac{i\slashed{x}}{2\pi^2 x^4}
       - \frac{m_q}{4\pi^2 x^2}
     \right]\delta^{ab}
     + \langle \chi_q^a(x)\,\bar{\chi}_q^b(0) \rangle
     \notag\\
  &\quad
     - \frac{ig_s}{32\pi^2}\,
       \frac{\slashed{x}\sigma_{\mu\nu}
             + \sigma_{\mu\nu}\slashed{x}}{x^2}\,
       G^{\mu\nu}_A(0)\,t_A^{ab}
     \notag\\
  &\quad
     + \frac{i}{3}\left[
         -\frac{\slashed{x}}{12}
         + \frac{1}{3}(u\cdot x)\slashed{u}
       \right]
       \langle u^{\mu}\Theta^f_{\mu\nu}u^{\nu}\rangle
       \delta^{ab}
     + \dots,
\end{align}
whereas the heavy-quark propagator  reads
\begin{align}\label{eq:heavyprop}
S_Q^{ab}(x)
  &= \frac{i}{(2\pi)^4}
     \int d^4k\,e^{-ik\cdot x}
     \Bigg\{
       \left(1-n_F\right)
       \frac{\delta_{ab}}
            {\slashed{k}-m_Q}
     \notag\\
  &\quad
       -\frac{g_sF_{\mu\nu}^{ab}(0)}{4}
       \frac{
       \sigma^{\mu\nu}(\slashed{k}+m_Q)
       +(\slashed{k}+m_Q)\sigma^{\mu\nu}
       }
       {(k^2-m_Q^2)^2}
     \notag\\
  &\quad
       +\frac{\pi^2}{3}
       \left\langle
       \frac{\alpha_s}{\pi}G^2
       \right\rangle
       \delta_{ab}\,
       m_Q
       \frac{k^2+m_Q\slashed{k}}
            {(k^2-m_Q^2)^4}
       +\dots
     \Bigg\}.
\end{align}

The statistical factor $(1-n_F)$ incorporates the Pauli-blocking effect in the hot and dense medium, where
\[
n_F=\left(e^{|p_0|/T}+1\right)^{-1}
\]
is the Fermi--Dirac distribution function. In the vacuum limit, corresponding to $T\to0$ and $\rho\to0$, Eqs.~(\ref{eq:lightprop}) and~(\ref{eq:heavyprop}) reduce to their respective vacuum propagators.

The coordinate-space expressions are subsequently transformed into
momentum space using the standard $D$-dimensional Fourier transformation
together with dimensional regularization techniques.
Since the intermediate algebraic manipulations are identical to those
presented in our previous analysis of heavy-light vector
$B$ mesons~\cite{Azizi:2026fnl},
they are not repeated here.

By equating the coefficients of the independent Lorentz structures in
the hadronic and QCD representations of the correlation function, one
obtains the coupled QCD sum rules,
\begin{eqnarray}\label{eq:SumRules}
-m_{D_{(s)}^*}^{*2}(T,\rho)\,
f_{D_{(s)}^*}^{*2}(T,\rho)\,
e^{-\kappa^2/M^2}
&=&
\Pi_{g_{\mu\nu}}^{\rm QCD}
(s_0,M^2,T,\rho),
\nonumber\\
f_{D_{(s)}^*}^{*2}(T,\rho)\,
e^{-\kappa^2/M^2}
&=&
\Pi_{p_\mu p_\nu}^{\rm QCD}
(s_0,M^2,T,\rho),
\nonumber\\
-\Sigma_v
f_{D_{(s)}^*}^{*2}(T,\rho)\,
e^{-\kappa^2/M^2}
&=&
\Pi_{p_\mu u_\nu}^{\rm QCD}
(s_0,M^2,T,\rho),
\nonumber\\
-\Sigma_v
f_{D_{(s)}^*}^{*2}(T,\rho)\,
e^{-\kappa^2/M^2}
&=&
\Pi_{p_\nu u_\mu}^{\rm QCD}
(s_0,M^2,T,\rho),
\nonumber\\
\Sigma_v^2
f_{D_{(s)}^*}^{*2}(T,\rho)\,
e^{-\kappa^2/M^2}
&=&
\Pi_{u_\mu u_\nu}^{\rm QCD}
(s_0,M^2,T,\rho).\nonumber\\
\end{eqnarray}

The coupled sum rules derived above form the basis of the numerical analysis. For completeness, the explicit QCD expression corresponding to the $p_\mu p_\nu$ Lorentz structure is presented in
Appendix~\ref{app:OPE}. It contains the perturbative contribution
together with the quark-, gluon-, and mixed quark--gluon-condensate
terms evaluated under finite-temperature and finite-density conditions.

%\FloatBarrier
%%%%%%%%%%%%%%%%%%%%%%%%%%%%%%%%%%%%%%%%%%%%%%%%%%%%%
\subsection{Temperature and Density Dependence of QCD Condensates}
%%%%%%%%%%%%%%%%%%%%%%%%%%%%%%%%%%%%%%%%%%%%%%%%%%%%%

The predictive power of in-medium QCD sum rules relies critically on a realistic description of the temperature- and density-dependent QCD condensates entering the operator product expansion. In the present work, the temperature- and density-dependent behavior of the light-quark and gluon condensates is based on the numerical results reported by Kumar \textit{et al.}~\cite{Kumar:2014}. Since analytic parametrizations were not provided in Ref.~\cite{Kumar:2014}, we employ the fitting functions extracted from their numerical results in our previous study of heavy-light vector mesons~\cite{Azizi:2026fnl}. These parametrizations consistently reproduce the simultaneous evolution of the light-quark and gluon condensates as functions of temperature and baryon density and are adopted here without further modification. As the fitting procedure and the resulting parametrizations have already been presented in detail in Ref.~\cite{Azizi:2026fnl}, only the final expressions required for the present analysis are summarized below. For the effective continuum threshold, we adopt the Hilbert-moment scaling prescription proposed by Dominguez, Loewe, and Rojas~\cite{Dominguez:2007}.

Based on the numerical results presented in Ref.~\cite{Kumar:2014}, the normalized light-quark condensate is parameterized as
\begin{equation}
\label{eq:qcKumar}
\frac{\langle\bar qq\rangle(T,\rho)}
{\langle\bar qq\rangle_0}
=
\exp\!\left[
-\frac{A\xi}{1+B\xi^{\kappa_1}}
+D\xi^\phi
\left(\frac{T}{T_c}\right)^\alpha
-
F
\left(\frac{T}{T_c}\right)^\beta
\right],
\end{equation}
where $\xi=\rho/\rho_0$, with
$\rho_0\simeq0.16~\mathrm{fm}^{-3}$ being the nuclear saturation density.
The corresponding fit parameters are
$A=0.5316$,
$B=0.1370$,
$\kappa_1=1.2262$,
$D=0.1375$,
$\phi=0.3374$,
$\alpha=1.2516$,
$F=0.0554$,
and
$\beta=20.0$.

\begin{figure*}[!t]
\centering
\begin{minipage}{0.48\textwidth}
\centering
\includegraphics[width=\linewidth]{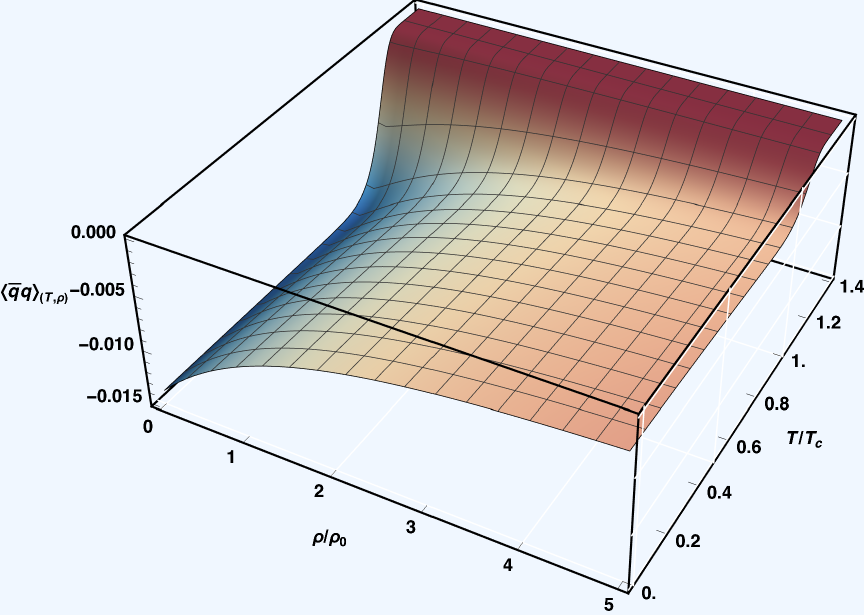}\\
\textbf{(a)}
\end{minipage}
\hfill
\begin{minipage}{0.48\textwidth}
\centering
\includegraphics[width=\linewidth]{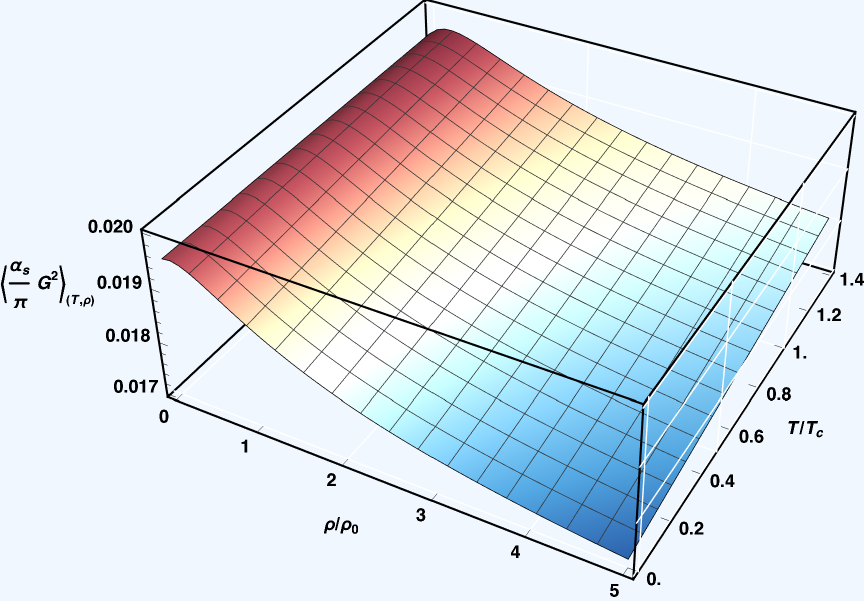}\\
\textbf{(b)}
\end{minipage}
\caption{\justifying
Temperature and density dependence of the QCD condensates employed in
the present analysis: (a) the light-quark condensate
$\langle\bar{q}q\rangle_{(T,\rho)}$ and (b) the gluon condensate
$\left\langle\frac{\alpha_s}{\pi}G^2\right\rangle_{(T,\rho)}$ as functions
of the normalized baryon density $\rho/\rho_0$ and the reduced
temperature $T/T_c$.}
\label{fig:Condensates}
\end{figure*}

The temperature and density dependence of the light-quark condensates
obtained from these expressions are shown in the left panel of
Fig.~\ref{fig:Condensates}. As the temperature and baryon density
increase, the magnitude of the condensate decreases monotonically,
reflecting the progressive melting of the chiral condensate in the
medium. Within the investigated thermodynamic domain, the suppression
becomes increasingly pronounced at high temperatures and densities,
indicating the gradual restoration of chiral symmetry. Since the
light-quark condensate constitutes the dominant nonperturbative input to
the QCD sum rules, its evolution largely determines the medium
dependence of the extracted masses and leptonic decay constants discussed in the
following sections.

The in-medium gluon condensate is described by the analytic
parametrization of Ref.~\cite{Kumar:2014},
\begin{align}
\label{eq:G2Kumar}
\left\langle
\frac{\alpha_s}{\pi}G^2
\right\rangle
(T,\rho)
&=
A_0
-
\frac{\xi}{T_c(1+\xi)^2}
\Big[
A_5T(1+\xi)
\nonumber\\
&
\qquad
+
T_c\Big((2+\xi)A_3+(1+\xi)A_4\Big)
\Big].
\end{align}
Here,
$A_0=\langle\frac{\alpha_s}{\pi}G^2\rangle_0$,
$A_3=-4.131\times10^{-3}~\mathrm{GeV}^4$,
$A_4=7.696\times10^{-3}~\mathrm{GeV}^4$,
and
$A_5=-8.530\times10^{-4}~\mathrm{GeV}^4$.

The corresponding temperature and density dependence of the gluon
condensate is shown in the right panel of
Fig.~\ref{fig:Condensates}. Compared with the light-quark condensate,
the gluon condensate exhibits a noticeably weaker dependence on baryon
density and remains nearly constant over a wide temperature interval.
Only as the temperature approaches the pseudocritical region does a
more pronounced suppression emerge, reflecting the increasing
modification of the nonperturbative gluonic vacuum structure. Although
its variation is considerably milder than that of
$\langle\bar{q}q\rangle(T,\rho)$, the gluon condensate constitutes an
essential nonperturbative input to the heavy-quark sector of the OPE and
therefore contributes to the medium dependence of the extracted hadronic
parameters.

Besides the leading quark and gluon condensates, the operator product
expansion contains several additional medium-dependent matrix elements,
including quark-number, mixed quark--gluon, and higher-dimensional
derivative condensates. Their temperature dependence is incorporated
within the same finite-temperature and finite-density framework adopted
for the leading condensates.

The light-quark propagator in Eq.~\eqref{eq:lightprop} contains the
thermal matrix element
$\langle u^\mu\Theta_{\mu\nu}^{f}u^\nu\rangle$, which represents the
fermionic contribution to the energy density of the medium. Following the
commonly adopted approximation of equal partitioning between the
fermionic and gluonic sectors~\cite{Veliev:2011kq}, we assume
\begin{equation}
\langle\Theta_{00}^{g}\rangle
=
\langle\Theta_{00}^{f}\rangle
=
\frac{1}{2}\langle\Theta_{00}\rangle,
\end{equation}
where the temperature dependence of the energy--momentum tensor is taken
from the lattice-QCD parametrization of
Refs.~\cite{Azizi:2015ona,Cheng:2007jq}.

The remaining condensates appearing in the sum rules are parametrized as
\begin{align}
\langle q^{\dagger}q\rangle_{(T,\rho)}
&=
\frac{3}{2}\rho, ~~
\langle s^{\dagger}s\rangle_{(T,\rho)}=0,
\\
\langle\bar qg_s\sigma Gq\rangle_{(T,\rho)}
&=
m_0^2\langle\bar qq\rangle_{(T,\rho)}
+
3\,\mathrm{GeV}^2\,\rho,
\\
\langle\bar sg_s\sigma Gs\rangle_{(T,\rho)}
&=
m_0^2\langle\bar ss\rangle_{(T,\rho)}
+
3y\,\mathrm{GeV}^2\,\rho,
\\
\langle q^{\dagger}g_s\sigma Gq\rangle_{(T,\rho)}
&=
-0.33\,\mathrm{GeV}^2\,\rho,
\\
\langle\bar qiD_0q\rangle_{(T,\rho)}
&=
0.18\,\rho,
\\
\langle\bar siD_0s\rangle_{(T,\rho)}
&=
m_s\langle\bar ss\rangle_{(T,\rho)}
+
0.02\,\rho,
\\
\langle\bar qiD_0iD_0q\rangle_{(T,\rho)}
&=
0.3\,\mathrm{GeV}^2\,\rho
-
\frac18
\langle\bar qg_s\sigma Gq\rangle_{(T,\rho)},
\end{align}
where the corresponding strange-quark operators are obtained using the
same scaling factor $y=0.05$.

Throughout this work, the mixed-condensate parameter is fixed at
$m_0^2=0.8~\mathrm{GeV}^2$, following
Ref.~\cite{Er:2022cxx} and the references therein. The quark masses are
taken from the latest PDG compilation,
$m_u=2.16\pm0.04~\mathrm{MeV}$,
$m_s=93.5\pm0.5~\mathrm{MeV}$, and
$m_c=1.2729\pm0.0027~\mathrm{GeV}$~\cite{PDG:2024}.

%\FloatBarrier
%%%%%%%%%%%%%%%%%%%%%%%%%%%%%%%%%%%%%%%%%%%%%%%%%%%%%
\subsection{Continuum Threshold and Borel Window}
%%%%%%%%%%%%%%%%%%%%%%%%%%%%%%%%%%%%%%%%%%%%%%%%%%%%%

To solve the coupled system of sum rules in
Eq.~\eqref{eq:SumRules}, the effective continuum threshold
$s_0(T,\rho)$ must be specified. Physically, this quantity defines
the separation between the ground-state pole and the onset of the
hadronic continuum. In vacuum, $s_0$ is conventionally chosen close
to the squared mass of the first excited state. In a hot and dense
medium, however, the continuum threshold is expected to decrease with
increasing temperature and baryon density, reflecting the gradual
modification of the hadronic spectrum.

The implementation of the medium-dependent continuum threshold within the QCD sum-rule framework was discussed in detail in our previous study of $B$-vector mesons~\cite{Azizi:2026fnl}. In the present work,
we employ the same prescription for the open-charm sector by adopting
the Hilbert-moment scaling relation proposed by Dominguez, Loewe, and
Rojas~\cite{Dominguez:2007},
\begin{align}
\label{eq:s0heavy}
\frac{s_0(T,\rho)}{s_0}
=
\frac{\langle\bar{q}q\rangle(T,\rho)}
{\langle\bar{q}q\rangle_0}
\left(
1-\frac{m_Q^2}{s_0}
\right)
+
\frac{m_Q^2}{s_0},
\end{align}
where $s_0$ denotes the vacuum continuum threshold and $m_Q$ is the
heavy-quark mass ($m_Q=m_c$ in the present analysis). Since the
in-medium continuum threshold is determined directly by the
light-quark condensate through Eq.~\eqref{eq:s0heavy}, its
temperature- and density-dependent evolution closely follows that of
$\langle\bar{q}q\rangle_{(T,\rho)}$. The pole dominance, OPE convergence and possible mild dependence of the physical observables on the auxiliary parameters leads to the interval
$(m_{D^*_{(s)}}+0.3)^2~\mathrm{GeV}^2 \le s_0 \le (m_{D^*_{(s)}}+0.5)^2~\mathrm{GeV}^2$
for the continuum threshold in vacuum.  A separate graphical representation
is therefore omitted to avoid redundant presentation of the same
physical information.

Another essential auxiliary parameter entering the QCD sum-rule
formalism is the Borel mass parameter $M^2$. Introduced through the
Borel transformation, it suppresses contributions from excited states
and the continuum on the hadronic side while simultaneously improving
the convergence of the operator product expansion on the QCD side. The
appropriate working interval, commonly referred to as the
\emph{Borel window}, is determined by simultaneously requiring
ground-state pole dominance and satisfactory OPE convergence.

To establish the numerical stability of the coupled sum rules in Eq.~\eqref{eq:SumRules}, we examine the dependence of the vacuum pole mass of the $D_s^{*\pm}$ meson on the Borel parameter $M^2$ for three representative continuum-threshold values, $s_0=5.8$, $6.3$, and $6.8~\mathrm{GeV}^2$. Since the charged $D_s^{*+}$ and $D_s^{*-}$ mesons are degenerate in vacuum and share the same PDG mass, only the $D_s^{*-}$ results are presented; the corresponding $D_s^{*+}$ curves are identical.

\begin{figure}[!t]
\centering
\includegraphics[width=0.95\linewidth]{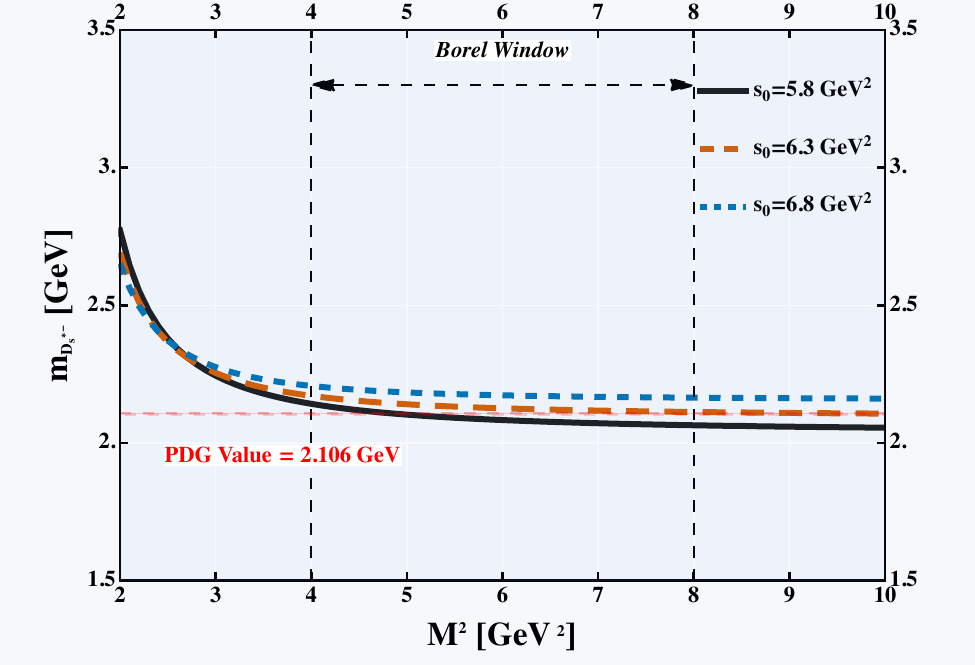}
\caption{\justifying
Vacuum pole mass of the $D_s^{*\pm}$ meson as a function of the Borel
mass parameter $M^2$ for three representative values of the continuum
threshold, $s_0=5.8$, $6.3$, and $6.8~\mathrm{GeV}^2$. The shaded region
indicates the working Borel window, determined from the simultaneous
requirements of OPE convergence and ground-state pole dominance. The
horizontal dashed line corresponds to the PDG value,
$m_{D_s^{*\pm}}=2.106~\mathrm{GeV}$~\cite{PDG:2024}.}
\label{fig:DSstarBorel}
\end{figure}

As shown in Fig.~\ref{fig:DSstarBorel}, the extracted pole mass exhibits
a stable plateau within the interval
$4.0~\mathrm{GeV}^2 \le M^2 \le 8.0~\mathrm{GeV}^2$, which is adopted as
the working Borel window throughout the numerical analysis common for all the states under study. For smaller
values of $M^2$, the prediction becomes increasingly sensitive to the
Borel parameter, indicating a deterioration of the OPE convergence.
Conversely, for larger values of $M^2$, the exponential suppression of
higher resonances becomes less effective, leading to an increasing
continuum contribution.

Within the adopted Borel window, the sum-rule prediction obtained with
the central continuum threshold,
$s_0=6.3~\mathrm{GeV}^2$, is in excellent agreement with the
experimental mass $m_{D_s^{*\pm}}=2.106~\mathrm{GeV}$ reported by the PDG~\cite{PDG:2024}. This agreement demonstrates the consistency of the adopted numerical
framework and provides confidence in the subsequent finite-temperature
and finite-density analysis.

%\FloatBarrier
%%%%%%%%%%%%%%%%%%%%%%%%%%%%%%%%%%%%%%%%%%%%%%%%%%%%%
%%%%%%%%%%%%%%%%%%%%%%%%%%%%%%%%%%%%%%%%%%%%%%%%%%%%%
\section{In-Medium Mass and Leptonic Decay Constant Shifts of the $D_s^{*\pm}$ and $D^{*\pm}$ Mesons}
\label{sec:Numeric}
%%%%%%%%%%%%%%%%%%%%%%%%%%%%%%%%%%%%%%%%%%%%%%%%%%%%%

Following the theoretical formulation presented in
Sec.~\ref{sec:Method}, we now turn to the numerical analysis of the
in-medium properties of the open-charm vector mesons
$D_s^{*}$ and $D^{*}$. The coupled sum rules derived in
Sec.~\ref{sec:Method} are numerically evaluated using the
temperature- and density-dependent QCD condensates together with the
corresponding medium-modified continuum threshold. The numerical
analysis aims to determine the in-medium masses,
$m_{D_{(s)}^{*}}^{*}(T,\rho)$, and leptonic decay constants,
$f_{D_{(s)}^{*}}^{*}(T,\rho)$, of both vector mesons.

Unless stated otherwise, all numerical results are obtained using the central values of the Borel parameter and the continuum threshold. Specifically, the Borel parameter is fixed at the midpoint of the working window, $4.0~\mathrm{GeV}^2 \le M^2 \le 8.0~\mathrm{GeV}^2$,
while the continuum threshold is taken at its central value as specified in Sec.~\ref{sec:Method}.

%\FloatBarrier
%%%%%%%%%%%%%%%%%%%%%%%%%%%%%%%%%%%%%%%%%%%%%%%%%%%%%
\subsection{In-medium Mass and Leptonic Decay Constant of the $D_s^{*-}$ Meson}
%%%%%%%%%%%%%%%%%%%%%%%%%%%%%%%%%%%%%%%%%%%%%%%%%%%%%
\begin{figure*}[!t]
\centering
\begin{minipage}{0.48\textwidth}
    \centering
    \includegraphics[width=\linewidth]{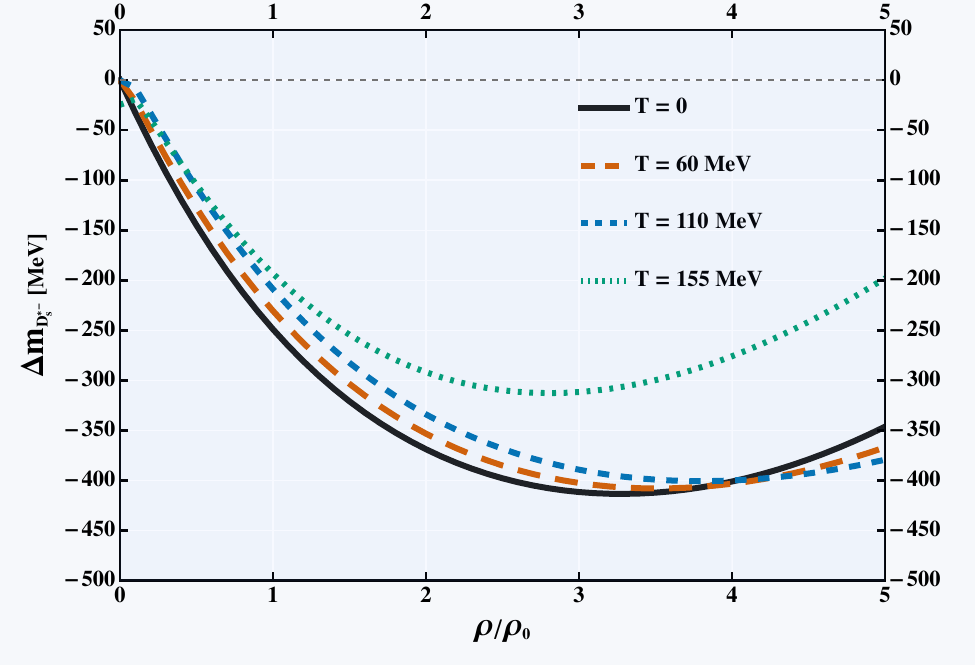}\\
    \textbf{(a)}
\end{minipage}
\hfill
\begin{minipage}{0.48\textwidth}
    \centering
    \includegraphics[width=\linewidth]{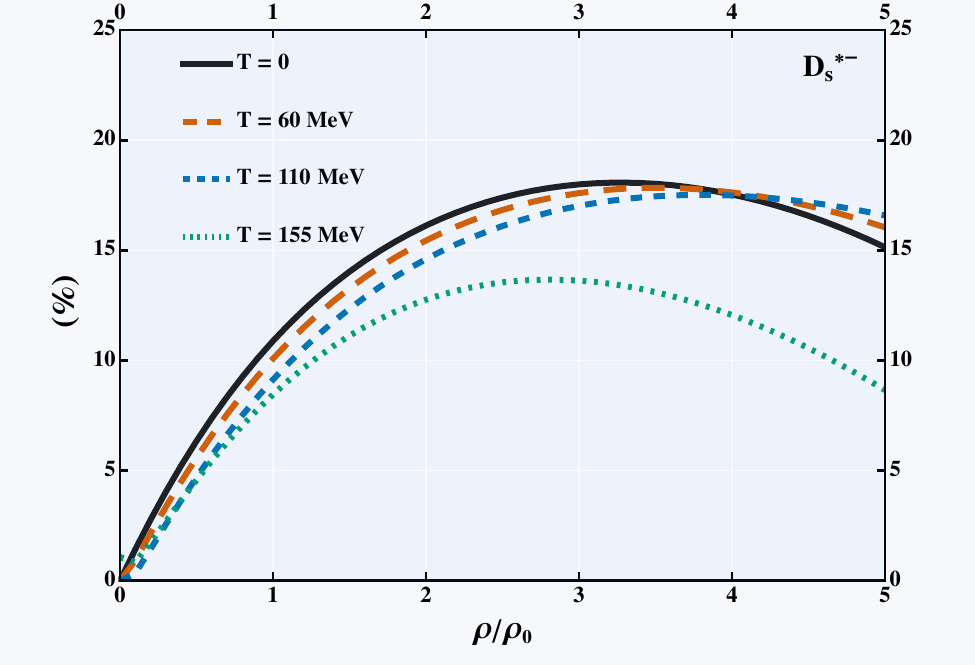}\\
    \textbf{(b)}
\end{minipage}
\caption{\justifying
(a) In-medium mass shift of the $D_s^{*-}$ meson,
$\Delta m_{D_s^{*-}}$, as a function of the normalized baryon density
$\rho/\rho_0$ at four representative temperatures:
$T=0$, $60~\mathrm{MeV}$,
$110~\mathrm{MeV}$, and
$155~\mathrm{MeV}$.
(b) Corresponding relative mass reduction,
$\left(1-\Delta m_{D_s^{*-}}/m_{D_s^{*-}}\right)\times100$,
as a function of $\rho/\rho_0$.}
\label{fig:DSstarMassRho}
\end{figure*}

The medium dependence of the $D_s^{*-}$ mass is presented in Fig.~\ref{fig:DSstarMassRho}. Panel~(a) displays the absolute mass shift,
defined as \[
\Delta m_{D_s^{*-}}
= m_{D_s^{*-}}^{*}-m_{D_s^{*-}},
\]
as a function of the normalized baryon density
$\rho/\rho_0$ for several temperatures.

The mass shift remains negative throughout the investigated
thermodynamic domain, indicating that the $D_s^{*-}$ meson experiences
an attractive in-medium interaction. At low temperatures
($T=0$--$110~\mathrm{MeV}$), the magnitude of the mass shift increases
rapidly with density and reaches its largest value around
$\rho/\rho_0\simeq3$--$3.5$, where
$\Delta m_{D_s^{*-}}\approx -(380$--$410)~\mathrm{MeV}$.
For higher densities, however, the curves exhibit a gradual recovery,
leading to a characteristic non-monotonic density dependence rather than
a continuous mass reduction.

At fixed baryon density, increasing the temperature systematically
reduces the magnitude of the negative mass shift. Consequently, the
$T=155~\mathrm{MeV}$ curve lies above the lower-temperature results over
the entire density range, with a minimum value of approximately
$-310~\mathrm{MeV}$ around $\rho/\rho_0\simeq 2.9$. These results indicate
that baryon density is the dominant source of the in-medium mass
modification, while finite temperature mainly moderates the
density-induced suppression. This behavior is consistent with the
temperature and density evolution of the condensates discussed in
Sec.~\ref{sec:Method}. A similar dominance of density effects has also
been reported in cold nuclear matter studies based on the
quark--meson-coupling model~\cite{Cobos-Martinez:2025iqg}, although those
calculations do not include thermal effects.

Another notable feature is the increasing separation of the
temperature-dependent curves with baryon density. At low densities
($\rho/\rho_0\lesssim1.5$), all four trajectories remain close to one
another, indicating relatively weak thermal effects. As the density
increases, the separation between the curves becomes progressively more
pronounced, demonstrating that thermal effects become increasingly
important in a dense medium through the combined temperature and density
dependence of the condensates entering the operator product expansion.

The same behavior is illustrated in panel~(b), where the relative mass
reduction,
$\left(1-\Delta m_{D_s^{*-}}/m_{D_s^{*-}}\right)\times100$,
is shown. Consistent with panel~(a), the relative reduction reaches its
maximum in the intermediate-density region
($\rho/\rho_0\simeq3$--$3.5$), attaining approximately
$17$--$18\%$ for
$T\le110~\mathrm{MeV}$ and about $14\%$ at
$T=155~\mathrm{MeV}$. At higher densities the relative suppression
decreases gradually, reflecting the same non-monotonic density
dependence observed for the absolute mass shift. The close correspondence
between panels~(a) and (b) confirms that the observed behavior is
independent of the normalization adopted and therefore represents a
robust prediction of the present QCD sum-rule analysis.

\begin{figure*}[!t]
\centering
\begin{minipage}{0.48\textwidth}
    \centering
    \includegraphics[width=\linewidth]{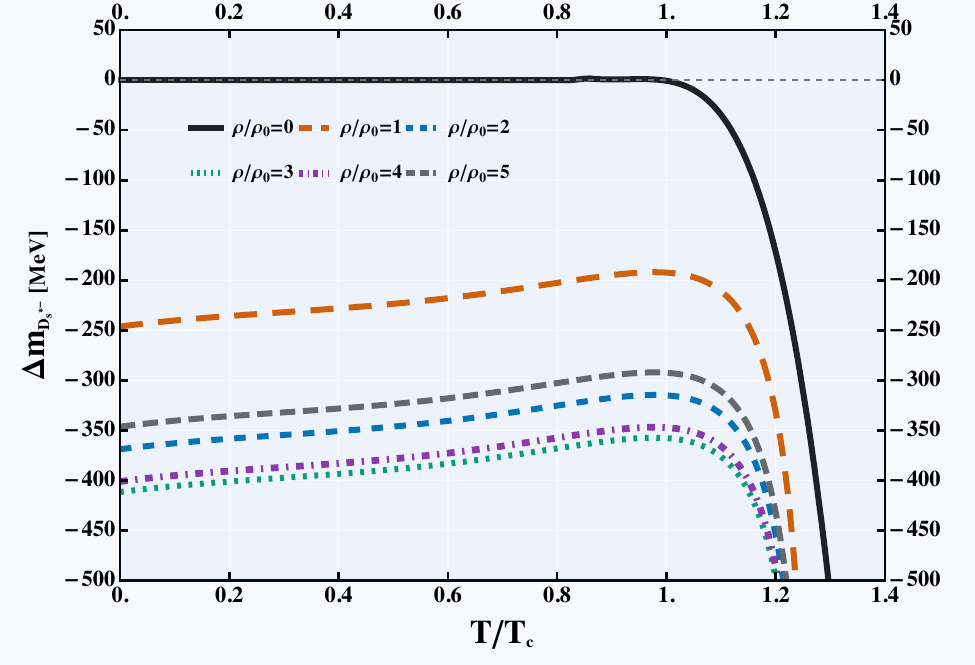}\\
    \textbf{(a)}
\end{minipage}
\hfill
\begin{minipage}{0.48\textwidth}
    \centering
    \includegraphics[width=\linewidth]{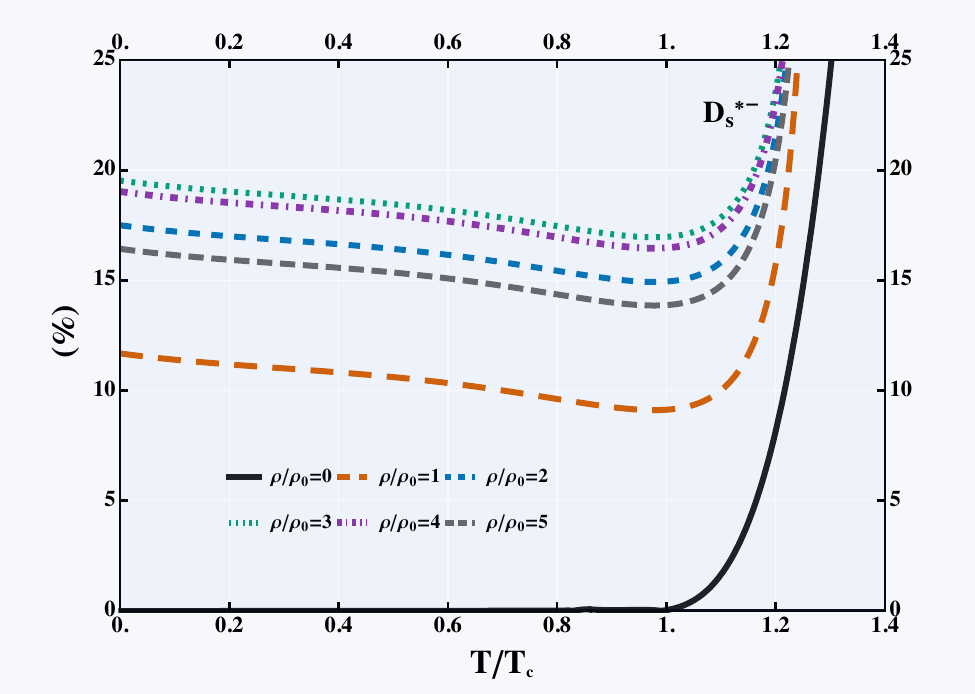}\\
    \textbf{(b)}
\end{minipage}
\caption{\justifying
(a) In-medium mass shift of the $D_s^{*-}$ meson,
$\Delta m_{D_s^{*-}}$, as a function of the reduced temperature
$T/T_c$ at fixed normalized baryon densities,
$\rho/\rho_0=0$, 1, 2, 3, 4, and 5.
(b) Corresponding relative mass reduction,
$\left(1-m_{D_s^{*-}}^{*}(T,\rho)/m_{D_s^{*-}}\right)\times100$,
as a function of $T/T_c$ for the same baryon densities.}
\label{fig:DSstarMassT}
\end{figure*}

Overall, the density dependence displayed in Fig.~\ref{fig:DSstarMassRho} demonstrates that baryon density provides the dominant contribution to the in-medium modification of the $D_s^{*-}$ mass, whereas finite temperature introduces comparatively smaller corrections. The complementary temperature evolution at fixed baryon densities is presented in Fig.~\ref{fig:DSstarMassT}. Panel~(a) shows the absolute mass shift, whereas panel~(b) presents the corresponding relative mass reduction with respect to the vacuum value.

\begin{figure*}[!t]
\centering
\begin{minipage}{0.48\textwidth}
    \centering
    \includegraphics[width=\linewidth]{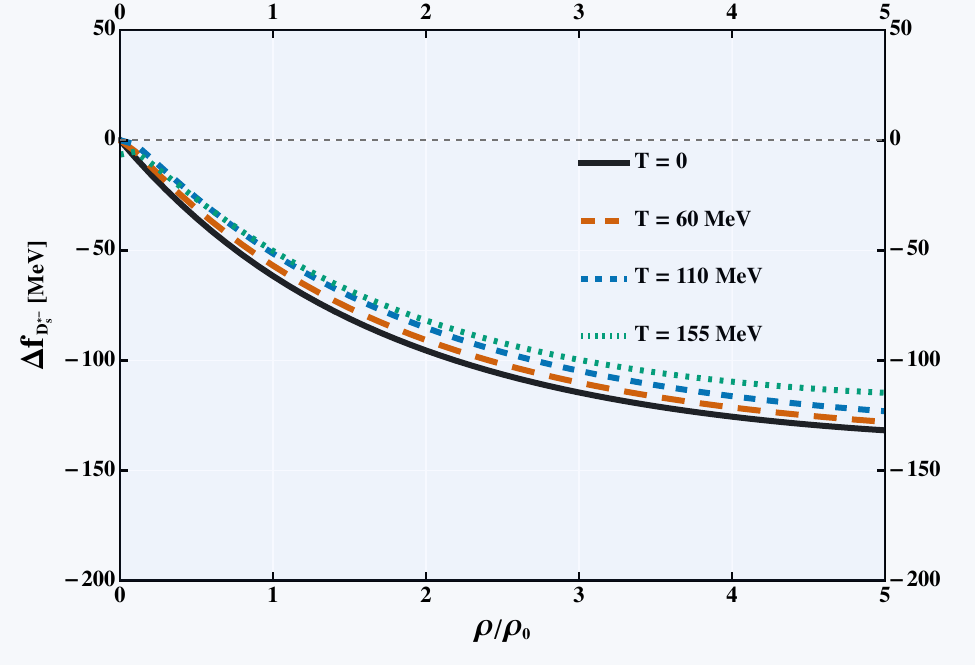}\\
    \textbf{(a)}
\end{minipage}
\hfill
\begin{minipage}{0.48\textwidth}
    \centering
    \includegraphics[width=\linewidth]{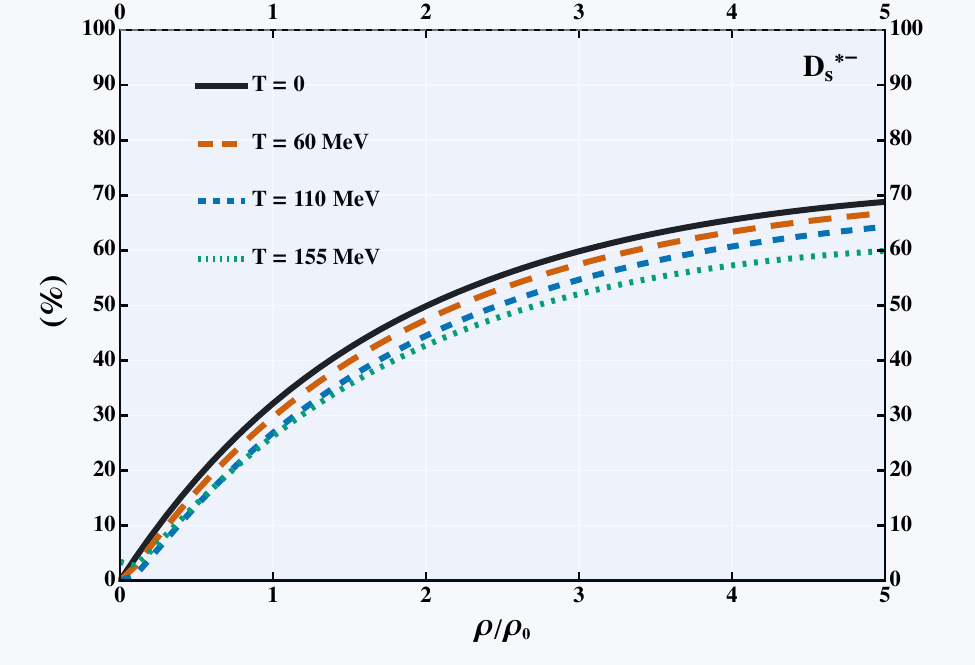}\\
    \textbf{(b)}
\end{minipage}
\caption{\justifying
Identical to Fig.~\ref{fig:DSstarMassRho}, but for the in-medium decay
constant of the $D_s^{*-}$ vector meson: (a) the decay-constant shift
$\Delta f_{D_s^{*-}}$ and (b) the corresponding relative change with
respect to its vacuum value, shown for the same temperatures and baryon
densities.}
\label{fig:DSstarfRho}
\end{figure*}

For all baryon densities considered, the mass remains nearly unchanged
throughout the low-temperature region up to approximately
$T/T_c\simeq1$. This indicates that thermal effects alone induce only
minor modifications of the relevant condensates in this regime.
Approaching the pseudocritical region, however, the magnitude of the
negative mass shift increases more rapidly, reflecting the accelerated
melting of the nonperturbative QCD condensates.

A second characteristic feature is the density dependence of the thermal
response. At $T=0$, the separation between neighboring density curves
decreases gradually as the density increases, particularly for
$\rho/\rho_0\gtrsim3$, indicating that the additional mass reduction
generated by successive density increments becomes progressively smaller
at high densities. This tendency persists over the entire temperature
range, where the high-density trajectories remain closely grouped,
whereas the low-density curves are more clearly separated.

The same behavior is reflected in the normalized results shown in
Fig.~\ref{fig:DSstarMassT}(b). The relative mass reduction remains
almost temperature independent up to approximately
$T/T_c\simeq1$ for all baryon densities. As the temperature approaches
the pseudocritical region, the suppression becomes progressively more
pronounced, with the effect being more visible at higher baryon
densities. This behavior indicates that thermal effects become
significant primarily near the pseudocritical temperature, where the
rapid evolution of the in-medium condensates enhances the overall
modification of the $D_s^{*-}$ mass.

The behavior of the $D_s^{*-}$ leptonic decay constant under hot and
dense nuclear matter conditions is presented in
Fig.~\ref{fig:DSstarfRho}. Panel~(a) shows the absolute shift,
$\Delta f_{D_s^{*-}}
=f_{D_s^{*-}}^{*}-f_{D_s^{*-}}$, while panel~(b) displays the
corresponding relative reduction with respect to the vacuum value.

For all temperatures considered, the leptonic decay constant decreases
monotonically with increasing baryon density. At the highest density,
$\rho/\rho_0=5$, the absolute reduction reaches approximately
$110$--$130~\mathrm{MeV}$, corresponding to a relative suppression of
about $60\%$--$70\%$. Unlike the mass, the leptonic decay constant shows
no turnaround over the investigated density range and instead decreases
monotonically with increasing baryon density.

The thermal evolution closely follows that observed for the mass.
Increasing the temperature at fixed baryon density weakens the
density-induced suppression, so that the
$T=155~\mathrm{MeV}$ curve lies systematically above the
lower-temperature curves throughout the entire density range. These
results indicate that baryon density provides the dominant contribution
to the in-medium modification of the $D_s^{*-}$ leptonic decay constant,
whereas finite temperature mainly reduces the magnitude of the
density-induced suppression without altering its monotonic dependence on
baryon density.

\begin{figure*}[!t]
\centering
\begin{minipage}{0.48\textwidth}
    \centering
    \includegraphics[width=\linewidth]{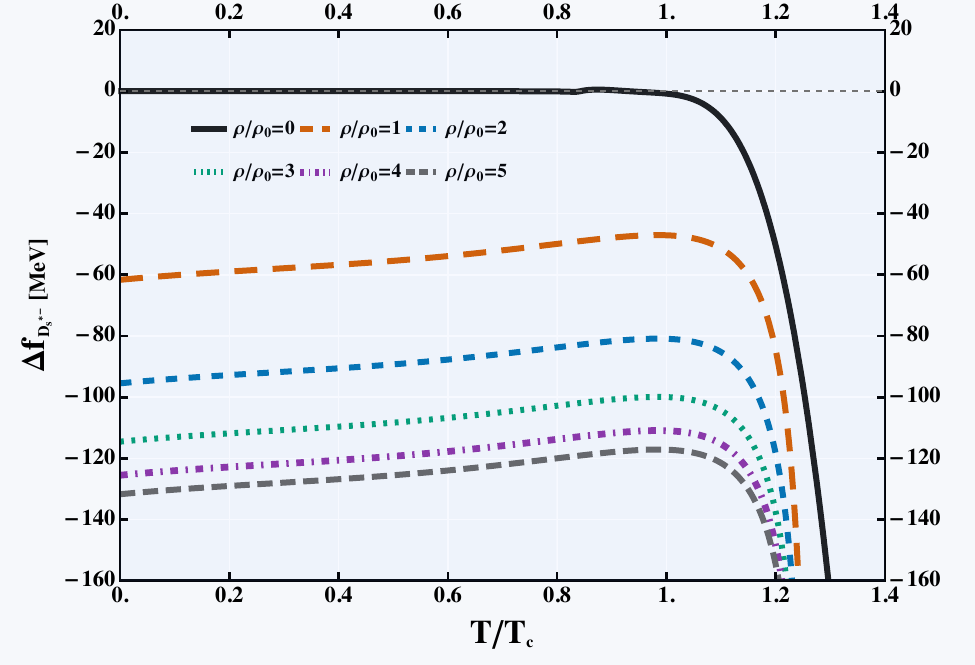}\\
    \textbf{(a)}
\end{minipage}
\hfill
\begin{minipage}{0.48\textwidth}
    \centering
    \includegraphics[width=\linewidth]{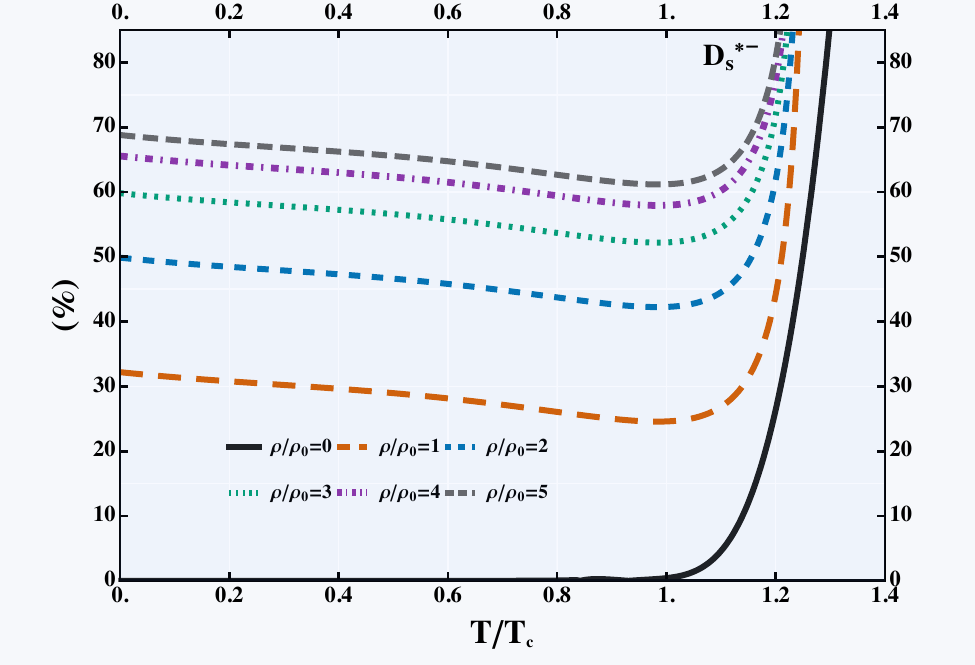}\\
    \textbf{(b)}
\end{minipage}
\caption{\justifying
Same as Fig.~\ref{fig:DSstarMassT}, but for the in-medium leptonic decay constant
of the $D_s^{*-}$ vector meson: (a) the decay-constant shift
$\Delta f_{D_s^{*-}}$ and (b) the percentage change relative to the
vacuum value under the same temperature and density conditions.}
\label{fig:DSstarfT}
\end{figure*}
The temperature dependence of the $D_s^{*-}$ leptonic decay constant at
fixed baryon densities is presented in
Fig.~\ref{fig:DSstarfT}, where panel~(a) shows the in-medium shift
$\Delta f_{D_s^{*-}}$ and panel~(b) the corresponding relative reduction.
For all baryon densities considered, the leptonic decay constant remains nearly
constant up to approximately $T/T_c\simeq1.0$. Consequently, both the
absolute shift and the relative suppression differ only marginally from
their $T=0$ values over this temperature range, indicating that thermal
effects below the pseudocritical region have only a limited influence on
the current--meson coupling compared with the modifications induced by
baryon density.

As the temperature approaches the pseudocritical region
($T/T_c\gtrsim1$), the leptonic decay constant decreases rapidly for all
baryon densities. This accelerated suppression follows the rapid thermal
evolution of the medium-dependent condensates near the chiral crossover.
Although the largest in-medium modifications remain associated with the
highest baryon densities, the temperature dependence becomes
significantly stronger in this region than at lower temperatures. Unlike
the mass, which exhibits a mild turnaround as a function of baryon
density, the leptonic decay constant shows no analogous behavior. Instead,
it decreases monotonically with temperature for all baryon densities
considered.

%\FloatBarrier
%%%%%%%%%%%%%%%%%%%%%%%%%%%%%%%%%%%%%%%%%%%%%%%%%%%%%
\subsection{In-medium Mass and leptonic decay constant of the $D^{*-}$ Vector Meson}
%%%%%%%%%%%%%%%%%%%%%%%%%%%%%%%%%%%%%%%%%%%%%%%%%%%%%
\begin{figure}[!t]
\centering
\begin{minipage}{0.48\textwidth}
    \centering
    \includegraphics[width=\linewidth]{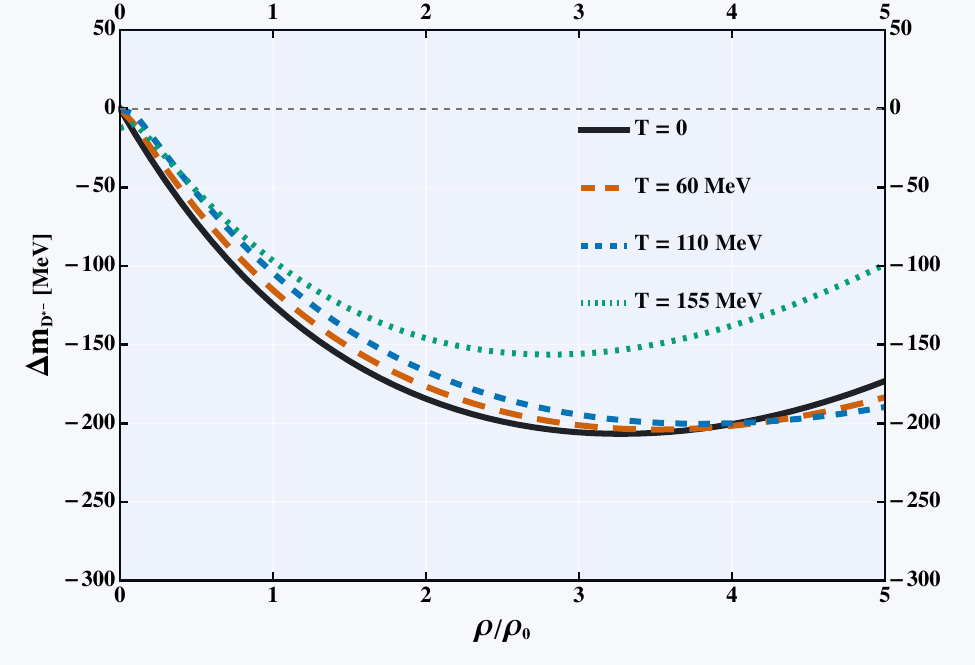}\\
    \textbf{(a)}
\end{minipage}
\hfill
\begin{minipage}{0.48\textwidth}
    \centering
    \includegraphics[width=\linewidth]{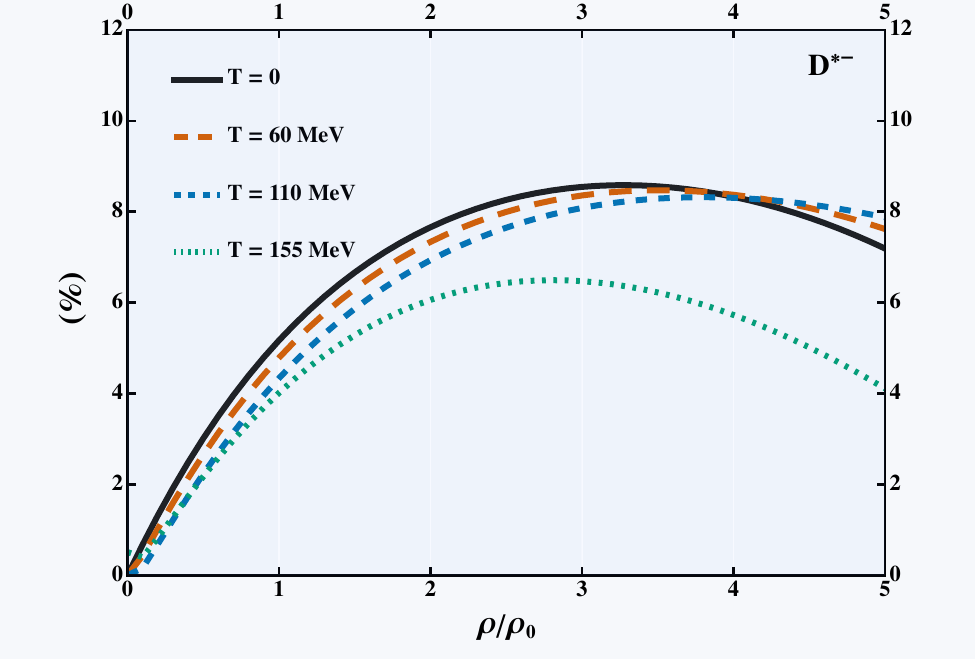}\\
    \textbf{(b)}
\end{minipage}
\caption{\justifying Density dependence of the in-medium $D^{*-}$ vector meson mass at different temperatures: (a) the mass shift $\Delta m_{D^{*-}}$ and (b) the corresponding percentage change relative to its vacuum value.}
\label{fig:DstarMassrho}
\end{figure}

The density dependence of the in-medium $D^{*-}$ mass at fixed
temperatures is presented in Fig.~\ref{fig:DstarMassrho}. Panel~(a)
shows the absolute mass shift,
$\Delta m_{D^{*-}}=m_{D^{*-}}^{*}-m_{D^{*-}}$, while panel~(b) displays
the corresponding relative reduction with respect to the vacuum mass.
For all temperatures considered, the mass shift becomes increasingly
negative as the baryon density increases, indicating a progressively
stronger in-medium mass reduction of the $D^{*-}$ meson. The suppression
reaches its maximum magnitude at
$\rho/\rho_0\simeq3$--3.5, after which the curves exhibit a gradual
recovery toward higher densities. The normalized results shown in
panel~(b) follow the same behavior, with the relative mass reduction
reaching its maximum in the intermediate-density region before
decreasing slightly at the highest densities.

The influence of temperature remains comparatively weak up to
$T=110~\mathrm{MeV}$, for which the corresponding trajectories are nearly
indistinguishable over the entire density range. In contrast, the
$T=155~\mathrm{MeV}$ curve is systematically shifted toward smaller
negative mass shifts, indicating that finite temperature partially
reduces the density-induced mass suppression. Compared with the $D_s^{*-}$ channel, the recovery beyond the intermediate-density minimum is less pronounced, leading to a smoother
high-density evolution of both the absolute and relative mass shifts.
Thus, although both vector mesons exhibit a similar non-monotonic density
dependence, the $D^{*-}$ state shows a weaker high-density recovery.
\begin{figure}[!t]
\centering
\begin{minipage}{0.48\textwidth}
    \centering
    \includegraphics[width=\linewidth]{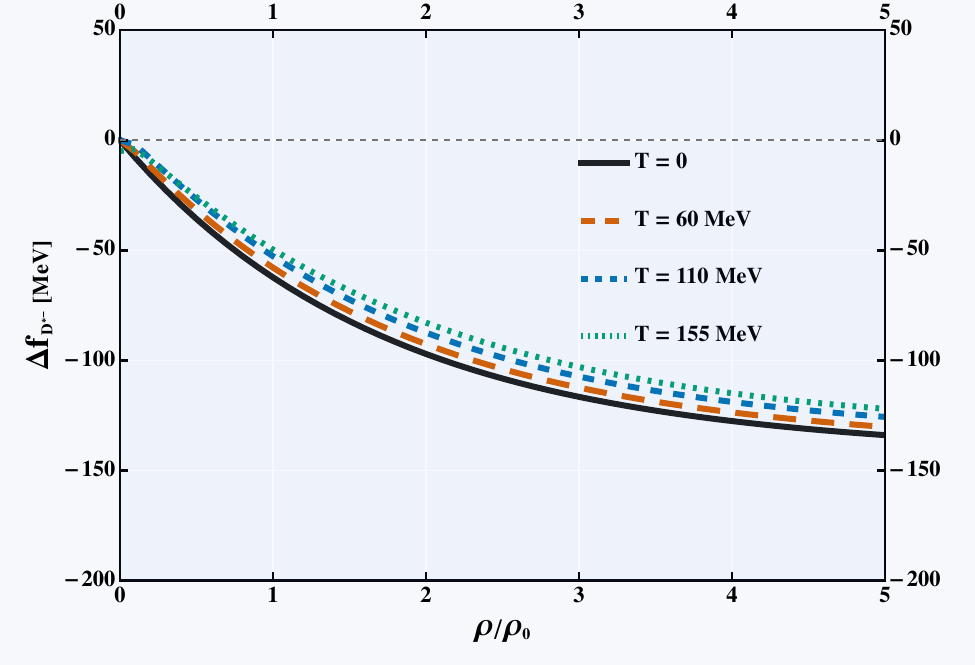}\\
    \textbf{(a)}
\end{minipage}
\hfill
\begin{minipage}{0.48\textwidth}
    \centering
    \includegraphics[width=\linewidth]{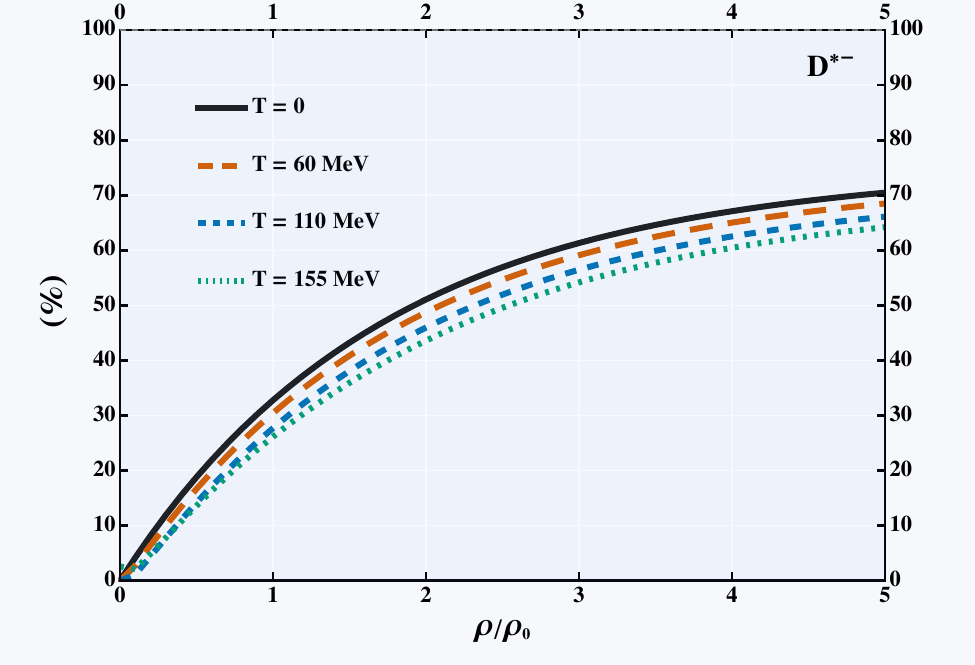}\\
    \textbf{(b)}
\end{minipage}
\caption{\justifying
Density dependence of the in-medium leptonic decay constant of the $D^{*-}$ vector
meson: (a) the leptonic decay constant shift $\Delta f_{D^{*-}}$ and (b) the
corresponding percentage change relative to the vacuum value, shown for
the same temperatures as in Fig.~\ref{fig:DstarMassrho}.}
\label{fig:Dstarfrho}
\end{figure}

To complement the pole-mass analysis,
Fig.~\ref{fig:Dstarfrho} presents the density dependence of the
$D^{*-}$ leptonic decay constant. Panel~(a) shows the in-medium shift
$\Delta f_{D^{*-}}$, while panel~(b) displays the corresponding
percentage reduction relative to the vacuum value.

For all temperatures considered, the leptonic decay constant decreases
monotonically with increasing baryon density, exhibiting no indication
of saturation or recovery over the investigated density range. This
behavior indicates a continuous reduction of the current--meson coupling
in dense nuclear matter and is qualitatively similar to that observed
for the $D_s^{*-}$ channel.

The in-medium evolution of the leptonic decay constant closely follows
that observed in the $D_s^{*-}$ channel. In both systems, increasing the
temperature systematically weakens the density-induced suppression while
preserving its monotonic dependence on baryon density. Consequently, the
highest-temperature curve remains consistently above the
lower-temperature curves throughout the investigated density range.
These results further confirm that baryon density is the primary source
of the in-medium modification of the $D^{*-}$ leptonic decay constant,
whereas finite temperature introduces comparatively smaller corrections.
%\FloatBarrier
%%%%%%%%%%%%%%%%%%%%%%%%%%%%%%%%%%%%%%%%%%%%%%%%%%%%%
\subsection{In-medium Antiparticle  $D_{(s)}^{*-}$-Particle $D_{(s)}^{*+}$ Splittings}
%%%%%%%%%%%%%%%%%%%%%%%%%%%%%%%%%%%%%%%%%%%%%%%%%%%%%

Having established the in-medium properties of the individual
$D_s^{*-}$ and $D^{*-}$ states, we now turn to the corresponding
particle--antiparticle splittings. Within the in-medium QCD sum-rule
framework, finite baryon density introduces the medium four-velocity
$u_\mu$, which breaks Lorentz invariance and generates an additional
odd component in the correlation function. As a consequence, the
particle and antiparticle channels are no longer degenerate, giving rise
to different in-medium self-energies and, consequently, distinct masses.
This mechanism was first established in QCD sum rules for
nucleons~\cite{Furnstahl:1992ux} and later extended to open-charm
mesons~\cite{Hilger2009}. In the present work, we investigate the
resulting particle--antiparticle splittings for both the masses and the
leptonic decay constants of the vector $D_{(s)}^{*\pm}$ mesons.

\begin{figure*}[!t]
\centering
\begin{minipage}{0.48\textwidth}
    \centering
    \includegraphics[width=\linewidth]{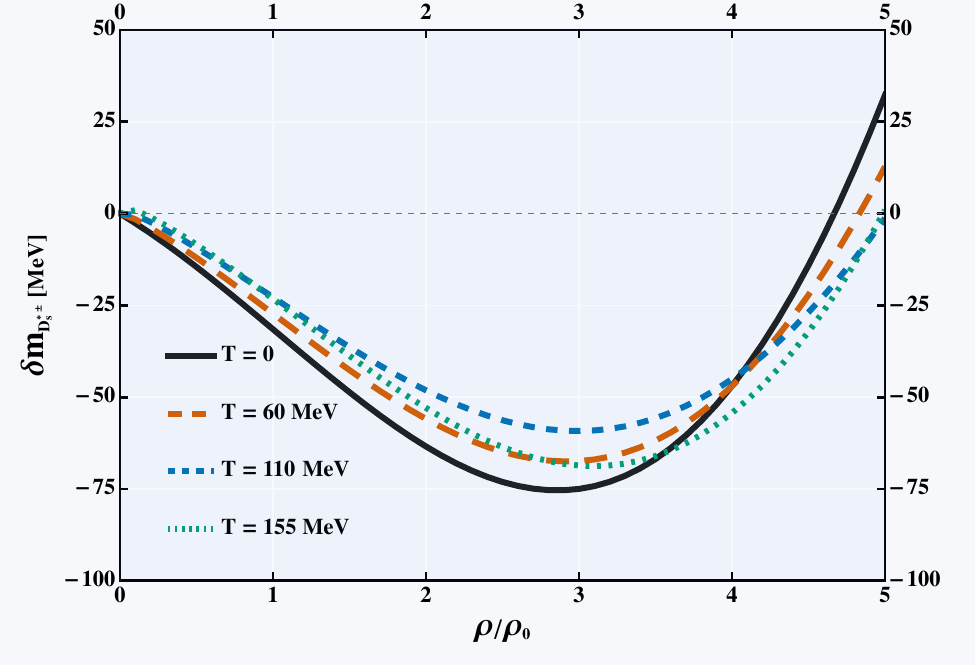}\\
    \textbf{(a)}
\end{minipage}
\hfill
\begin{minipage}{0.48\textwidth}
    \centering
    \includegraphics[width=\linewidth]{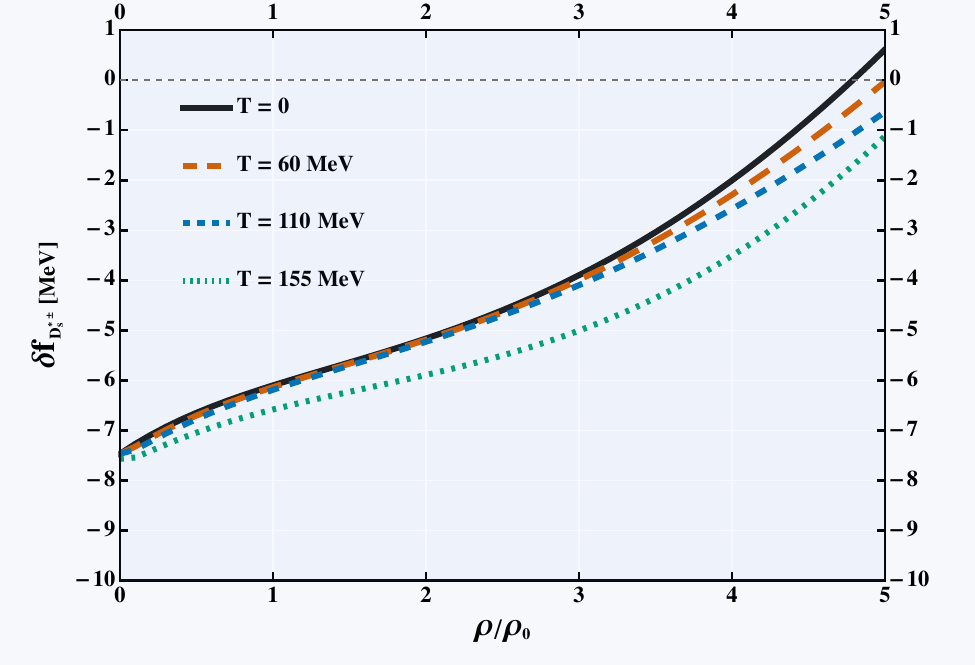}\\
    \textbf{(b)}
\end{minipage}
\caption{\justifying
Density dependence of the antiparticle-particle splitting in the
$D_s^{*\pm}$ system at different temperatures:
(a) the in-medium mass difference
$\delta m = m_{D_s^{*-}}^{*}-m_{D_s^{*+}}^{*}$ and
(b) the in-medium decay-constant difference
$\delta f = f_{D_s^{*-}}^{*}-f_{D_s^{*+}}^{*}$,
plotted under identical temperature and density conditions.}
\label{fig:DSstarMassMinusPlus}
\end{figure*}

The in-medium particle-antiparticle splittings of the
$D_s^{*\pm}$ vector mesons are presented in
Fig.~\ref{fig:DSstarMassMinusPlus}. Panel~(a) shows the mass difference,
defined as
$\delta m=m_{D_s^{*-}}^{*}-m_{D_s^{*+}}^{*}$,
while panel~(b) displays the corresponding decay-constant difference,
$\delta f=f_{D_s^{*-}}^{*}-f_{D_s^{*+}}^{*}$, as functions of the
normalized baryon density.

As shown in Fig.~\ref{fig:DSstarMassMinusPlus}(a), the mass splitting
vanishes in vacuum and becomes increasingly negative with increasing
baryon density, demonstrating that the $D_s^{*-}$ state undergoes a
stronger in-medium mass reduction than its antiparticle. For
$T\le110~\mathrm{MeV}$, the splitting reaches its largest magnitude at
$\rho/\rho_0\simeq2.8$--$3.0$, with the position of the minimum remaining
essentially unchanged. A different behavior is observed at
$T=155~\mathrm{MeV}$, where both the location and the depth of the
minimum deviate from the lower-temperature trend. In particular, the
minimum shifts to $\rho/\rho_0\simeq3.2$, and its magnitude becomes
slightly larger than those obtained at $T=60$ and
$110~\mathrm{MeV}$. This indicates that the temperature dependence of the
particle--antiparticle mass splitting is no longer monotonic as the
system approaches the pseudocritical region. Nevertheless, away from the
minimum, increasing temperature generally reduces the magnitude of the
splitting without removing it over the density range considered.

A different behavior is observed for the decay-constant splitting shown
in Fig.~\ref{fig:DSstarMassMinusPlus}(b). In contrast to the mass
splitting, $\delta f$ decreases monotonically in magnitude with
increasing baryon density and exhibits no intermediate minimum. The
largest splitting is obtained in vacuum, after which the difference
between the $D_s^{*-}$ and $D_s^{*+}$ leptonic decay constants gradually
diminishes, approaching zero at the highest densities considered.

The temperature dependence of the decay-constant splitting is also
qualitatively different from that of the mass splitting. Whereas the
curves for $T\le110~\mathrm{MeV}$ remain relatively close to one another,
the $T=155~\mathrm{MeV}$ trajectory is shifted toward more negative
values over the entire density range. As a result, the decay-constant
splitting becomes slightly larger at the highest temperature considered.
In addition, the separation between the temperature-dependent curves
increases gradually with baryon density, indicating that thermal effects
become more visible in dense matter, although the overall splitting
remains considerably smaller than the density-induced modification of the
individual leptonic decay constants.
\begin{figure}[!t]
\centering
\begin{minipage}{0.48\textwidth}
    \centering
    \includegraphics[width=\linewidth]{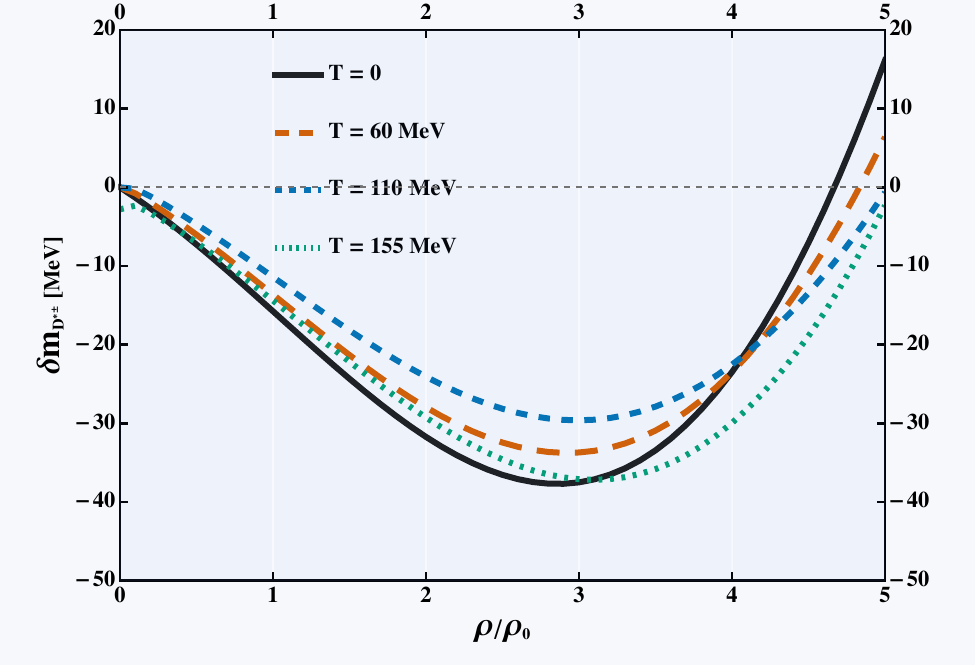}\\
    \textbf{(a)}
\end{minipage}
\hfill
\begin{minipage}{0.48\textwidth}
    \centering
    \includegraphics[width=\linewidth]{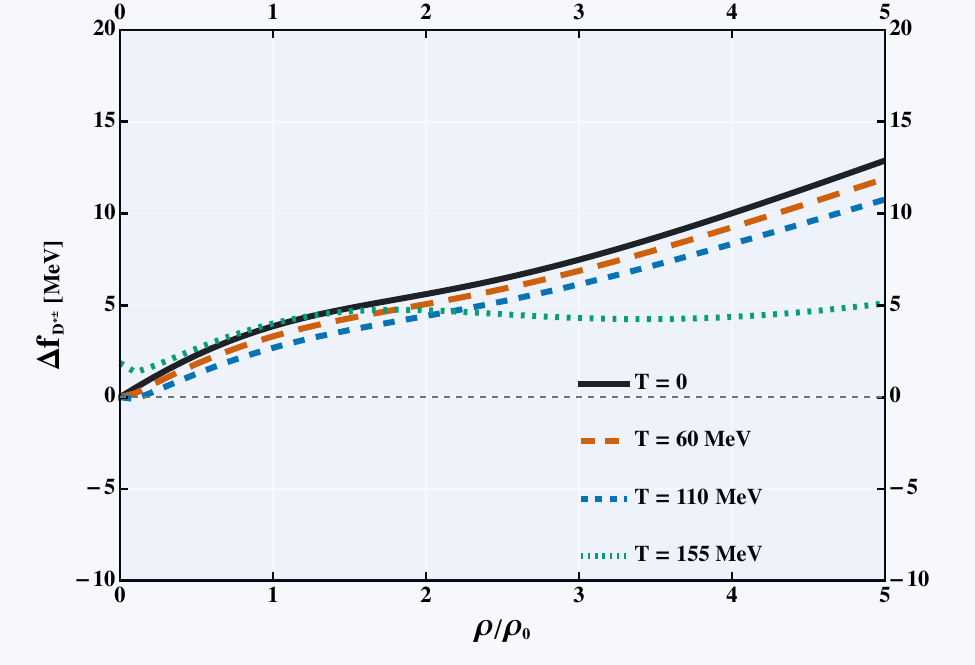}\\
    \textbf{(b)}
\end{minipage}
\caption{\justifying Density dependence of the splitting between the
$D^{*-}$ and $D^{*+}$ states at different temperatures:
(a) the mass difference
$\delta m = m_{D^{*-}} - m_{D^{*+}}$
and (b) the leptonic decay constant difference
$\delta f = f_{D^{*-}} - f_{D^{*+}}$,
under the same medium conditions.}
\label{fig:Dstarminusplus}
\end{figure}

\begin{table*}[!t]
\centering
\caption{\justifying Summary of the principal in-medium properties of the strange
$D_s^{*-}$ and non-strange $D^{*-}$ vector mesons obtained in the present
QCD sum-rule analysis. For each observable, the temperature (in MeV) and
normalized baryon density $(\rho/\rho_0)$ at which the extremum occurs
are also listed.}
\label{tab:SummaryComparison}
\renewcommand{\arraystretch}{1.2}
\begin{tabular}{lcc}
\toprule
\textbf{Observable} &
\textbf{$D_s^{*-}$} &
\textbf{$D^{*-}$} \\
\midrule

Maximum mass shift,
$|\Delta m|_{\rm max}$ (MeV)
&
\shortstack[c]{$413.2$\\$(T=0,\ \rho/\rho_0=3.3)$}
&
\shortstack[c]{$206.5$\\$(T=0,\ \rho/\rho_0=3.15)$}
\\[2mm]

Maximum decay-constant reduction,
$|\Delta f|_{\rm max}$ (MeV)
&
\shortstack[c]{$131.7$\\$(T=0,\ \rho/\rho_0=5.0)$}
&
\shortstack[c]{$133.8$\\$(T=0,\ \rho/\rho_0=5.0)$}
\\[2mm]

Maximum relative mass reduction (\%)
&
\shortstack[c]{$18.1$\\$(T=0,\ \rho/\rho_0=3.3)$}
&
\shortstack[c]{$8.5$\\$(T=0,\ \rho/\rho_0=3.15)$}
\\[2mm]

Maximum relative decay-constant reduction (\%)
&
\shortstack[c]{$68.8$\\$(T=0,\ \rho/\rho_0=5.0)$}
&
\shortstack[c]{$70.5$\\$(T=0,\ \rho/\rho_0=5.0)$}
\\[2mm]

Maximum mass splitting,
$|\delta m|_{\rm max}$ (MeV)
&
\shortstack[c]{$75.3$\\$(T=0,\ \rho/\rho_0=2.9)$}
&
\shortstack[c]{$37.7$\\$(T=0,\ \rho/\rho_0=2.9)$}
\\[2mm]

Maximum decay-constant splitting,
$|\delta f|_{\rm max}$ (MeV)
&
\shortstack[c]{$7.6$\\$(T=155,\ \rho/\rho_0=0)$}
&
\shortstack[c]{$1.8$\\$(T=155,\ \rho/\rho_0=0)$}
\\

\addlinespace

Density dependence of the mass
&
Non-monotonic
&
Non-monotonic
\\[2mm]

Density dependence of the decay constant
&
Monotonic suppression
&
Monotonic suppression
\\[2mm]

Dominant medium effect
&
Baryon density
&
Baryon density
\\[2mm]

Thermal response
&
\shortstack[c]{Secondary; moderates\\density effects}
&
\shortstack[c]{Secondary; moderates\\density effects}
\\

\bottomrule
\end{tabular}
\end{table*}

As shown in Fig.~\ref{fig:Dstarminusplus}(a), the mass splitting
$\delta m_{D^{*\pm}}$ vanishes in vacuum and becomes increasingly
negative with increasing baryon density, reaching its largest magnitude
around $\rho/\rho_0\simeq3$. Beyond this density, the splitting
gradually decreases in magnitude and eventually changes sign at the
highest densities for the $T=0$ and $T=60~\mathrm{MeV}$ curves,
indicating a partial restoration of the particle--antiparticle mass
degeneracy; the $T=110$ and $155~\mathrm{MeV}$ curves remain negative
throughout the density range considered. For $T\le110~\mathrm{MeV}$,
increasing temperature reduces the depth of the minimum. However, this
trend is no longer maintained at $T=155~\mathrm{MeV}$, where the minimum
becomes nearly as deep as the vacuum-density $T=0$ result, indicating a
departure from the otherwise monotonic temperature dependence.

A qualitatively different behavior is observed for the decay-constant
splitting shown in Fig.~\ref{fig:Dstarminusplus}(b). For
$T\le110~\mathrm{MeV}$, $\delta f_{D^{*\pm}}$ increases monotonically
with baryon density while remaining positive throughout the investigated
region. In contrast, the $T=155~\mathrm{MeV}$ trajectory rises rapidly at
low density but then saturates, exhibiting a much weaker density
dependence and approaching a broad plateau over the intermediate- and
high-density region. Consequently, the decay-constant splitting shows a
much weaker sensitivity to increasing baryon density at the highest
temperature than at lower temperatures. Notably, this monotonic growth of
$\delta f_{D^{*\pm}}$ with density stands in clear contrast to the
behavior found for $\delta f_{D_s^{*\pm}}$, which instead decreases in
magnitude toward higher densities -- highlighting a genuine
flavor-dependent difference between the non-strange and strange
open-charm vector channels.

%%%%%%%%%%%%%%%%%%%%%%%%%%%%%%%%%%%%%%%%%%%%%%%%%%%%%
\subsection{Overall Comparison}
%%%%%%%%%%%%%%%%%%%%%%%%%%%%%%%%%%%%%%%%%%%%%%%%%%%%%
Table~\ref{tab:SummaryComparison} provides a consolidated comparison of
the principal in-medium properties of the $D_s^{*-}$ and $D^{*-}$ vector
mesons obtained in the present analysis. In addition to the extrema of
the mass and decay-constant modifications, it summarizes the dominant
medium dependence and the characteristic thermodynamic behavior of both
systems, thereby facilitating a direct comparison between the strange
and non-strange open-charm sectors.

As summarized in Table~\ref{tab:SummaryComparison}, the two vector
mesons exhibit remarkably similar qualitative responses to hot and dense
nuclear matter, indicating that the underlying mechanism governing their
in-medium evolution is largely common to both channels. In both cases,
baryon density provides the dominant source of the medium modification,
whereas finite temperature mainly alters the magnitude of these effects
as the system approaches the pseudocritical region.

Despite these common features, several quantitative differences emerge.
The $D_s^{*-}$ meson undergoes systematically larger in-medium mass
shifts than the $D^{*-}$ state, particularly in the vicinity of the
intermediate-density minimum where the largest deviations from the
vacuum masses are observed. By contrast, the leptonic decay constants of the two
channels exhibit comparable magnitudes of suppression, differing mainly
in overall strength rather than in their qualitative density dependence.

A further distinction is found in the particle--antiparticle mass
splitting: while its magnitude is substantially larger in the strange
channel than in the non-strange one, both $D_s^{*\pm}$ and $D^{*\pm}$
splittings follow a qualitatively similar pattern, growing in magnitude
up to $\rho/\rho_0\simeq3$ and subsequently changing sign at the highest
densities considered for the lower-temperature curves -- indicating a
partial restoration of the particle--antiparticle mass degeneracy in
both channels. These observations demonstrate that the strange quark
primarily influences the quantitative strength of the in-medium
modifications and splittings, whereas the overall thermodynamic
evolution of the open-charm vector mesons remains qualitatively similar
in the two systems.

% \FloatBarrier
%%%%%%%%%%%%%%%%%%%%%%%%%%%%%%%%%%%%%%%%%%%%%%%%%%%%%
\section{Conclusion}
%%%%%%%%%%%%%%%%%%%%%%%%%%%%%%%%%%%%%%%%%%%%%%%%%%%%%
In this work, we have investigated the in-medium properties of the
strange $D_s^{*\pm}$ and non-strange $D^{*\pm}$ vector mesons within the
framework of finite-temperature and finite-density QCD sum rules. Using
temperature- and density-dependent QCD condensates, we have determined
their in-medium masses, leptonic decay constants, and
particle--antiparticle mass and decay-constant splittings over a broad
thermodynamic domain relevant to hot and dense hadronic matter.

The results demonstrate that the open-charm vector sector responds to
the nuclear medium primarily through the density dependence of the
medium-modified QCD condensates, while finite temperature acts mainly as
a secondary correction to this behavior. The observed non-monotonic
density dependence of the masses, together with the finite
particle--antiparticle splittings generated in the baryonic medium, is
consistent with the medium evolution of the QCD condensates associated
with the partial restoration of chiral symmetry. The comparison between
the strange and non-strange channels further indicates that this
underlying mechanism is common to both systems, with the strange-quark
content primarily affecting the magnitude of the medium modifications
rather than their qualitative behavior.

The predictions presented here provide quantitative benchmarks for the
behavior of open-charm vector mesons in hot and dense strongly
interacting matter. They may serve as theoretical input for future
studies of charmed hadrons in nuclear matter and for the interpretation
of open-charm observables at forthcoming heavy-ion facilities such as
FAIR-CBM, NICA, and J-PARC, where medium modifications of heavy-flavor
hadrons are expected to play an important role. A natural extension of
this work would incorporate finite-width effects and a more complete
treatment of the four-quark condensates, which remain the dominant
source of theoretical uncertainty in the present approach.

%%%%%%%%%%%%%%%%%%%%%%%%%%%%%%%%%%%%%%%%%%%%%%%%%%%%%

% \FloatBarrier
 %%%%%%%%%%%%%%%%%%%%%%%%%%%%%%%%%%%%%%%%%%%%%%%%%%%%%
\begin{acknowledgments}

K. A. thanks the Iran National Science Foundation
(INSF) for partial financial support provided under the
Elite Grant No. 40405095.

\end{acknowledgments}

\section*{Declaration of AI-assisted language editing}

The authors have used AI tools, only for language editing and improving the readability of the manuscript. The authors have reviewed and edited the manuscript and take full responsibility for its scientific content.

%%%%%%%%%%%%%%%%%%%%%%%%%%%%%%%%%%%%%%%%%%%%%%%%%%%%%

%\FloatBarrier
\appendix
\section{Analytical Expressions for the $p_{\mu} p_{\nu}$ Structure}
\label{app:OPE}
%%%%%%%%%%%%%%%%%%%%%%%%%%%%%%%%%%%%%%%%%%%%%%%%%%%%%
For completeness, this appendix collects the explicit analytical
expression for the QCD side of the correlation function corresponding
to the Lorentz structure $p_{\mu}p_{\nu}$, denoted by
$\Pi^{\mathrm{QCD}}_{p_{\mu} p_{\nu}}(s_0,M^2,T,\rho)$, for the
open-charm vector meson $D_s^{*-}$ considered in the present work. The
expression incorporates the combined effects of finite temperature and
baryon density through the medium-dependent condensates entering the
operator product expansion.
\begin{widetext}
\begin{eqnarray}
\Pi^{\mathrm{QCD}}_{p_{\mu} p_{\nu}} (s_0,M^2,T,\rho) &=& \frac{3(1-n_F)^2}{4\pi^2} \int_{(m_c+m_s)^2}^{s_0} ds\,e^{-s/M^2} \int_{0}^{1} dz\, \,z(1-z)\, \Theta\!\left[L(s,z)\right] \nonumber \\
&+&\frac{4\,e^{-m_c^2/M^2}\,(n_F-1)\,\langle\theta_{00}^f\rangle}
{3M^2} + \frac{1}{12M^4} \langle \frac{\alpha_s}{\pi} G^2\rangle 
\int_0^1 dz\, \frac{ e^{-\frac{m_c^2}{M^2(1-z)}} \left[ M^2(1-z)^2 +(n_F-1)m_c^2z \right] } {(1-z)^2}
\nonumber\\
&-&\frac{4\,(n_F-1)} {3M^4}\, e^{-m_c^2/M^2} \left( m_s M^2 \langle \bar{s}\,s\rangle -2p_0\langle s^\dagger iD_0 iD_0 s\rangle -M^2\langle s^\dagger iD_0 s\rangle \right).
\end{eqnarray}
\end{widetext}
Here $\Theta(x)$ denotes the unit step (Heaviside) function, which
restricts the integration to the kinematically allowed region, and
$p_0$ denotes the energy of the meson in the rest frame of the medium.
The quantity $L(s,z)$ appearing in its argument is defined by
\begin{equation}
L(s,z)=-m_c^2z+s\,z(1-z),
\label{eq:L}
\end{equation}
where $s$ is the dispersion (invariant-mass-squared) variable and
$z\in[0,1]$ is the corresponding Feynman parameter.

%%%%%%%%%%%%%%%%%%%%%%%%%%%%%%%%%%%%%%%%%%%%%%%%%%%%%

% \FloatBarrier


\begin{thebibliography}{99}

%\cite{Fukushima2010ThePD}
\bibitem{Shuryak}
E. V. Shuryak,
"Quantum chromodynamics and the theory of superdense matter'',
\href{https://doi.org/10.1016/0370-1573(80)90105-2}{Physics Reports \textbf{61}, 71-158 (1980)}.


%\cite{Fukushima2010ThePD}
\bibitem{Fukushima2010ThePD}
K. Fukushima and T. Hatsuda,
"The phase diagram of dense QCD'',
\href{https://api.semanticscholar.org/CorpusID:119108371}{Reports on Progress in Physics \textbf{74}, 014001 (2010)}.

\bibitem{Karsch2002}
F.~Karsch,
"Lattice QCD at High Temperature and Density",
\href{https://inspirehep.net/literature/559076}{Lect.\ Notes Phys.\ \textbf{583}, 209 (2002)}.

\bibitem{HatsudaLee1992}
T. Hatsuda and S. H. Lee,
"QCD sum rules for vector mesons in the nuclear medium",
\href{https://link.aps.org/doi/10.1103/PhysRevC.46.R34}{Phys. Rev. C \textbf{46}, R34 (1992)}.

\bibitem{RappWambach2000}
R. Rapp and J. Wambach,
"Chiral Symmetry Restoration and Dileptons in Relativistic Heavy-Ion Collisions",
\href{https://inspirehep.net/literature/506479}{Adv. Nucl. Phys. \textbf{25}, 1 (2000)}.

%\cite{Hayashigaki:2000es}
\bibitem{Hayashigaki:2000es}
A.~Hayashigaki,
``Mass modification of D meson at finite density in QCD sum rule,''
\href{https://www.sciencedirect.com/science/article/pii/S0370269300007607?via%3Dihub}{Phys. Lett. B \textbf{487} , 96-103 (2000)}.

\bibitem{Arsene2005}
I.~Arsene \emph{et al.} [BRAHMS Collaboration],
"Quark Gluon Plasma and Color Glass Condensate at RHIC? The Perspective from the BRAHMS Experiment",
\href{https://www.sciencedirect.com/science/article/pii/S0375947405002770}{Nucl.\ Phys.\ A \textbf{757}, 1 (2005)}.

\bibitem{Adams2005}
J.~Adams \emph{et al.} [STAR Collaboration],
"Experimental and theoretical challenges in the search for the quark gluon plasma: The STAR Collaboration's critical assessment of the evidence from RHIC collisions",
\href{https://www.sciencedirect.com/science/article/pii/S0375947405005294?via%3Dihub}{Nucl.\ Phys.\ A \textbf{757}, 102-183 (2005)}.

%\cite{ALICE:2010khr}
\bibitem{ALICE:2010khr}
K.~Aamodt \textit{et al.} [ALICE],
``Charged-particle multiplicity density at mid-rapidity in central Pb-Pb collisions at $\sqrt{s_{NN}} = 2.76$ TeV,''
\href{https://doi.org/10.1103/PhysRevLett.105.252301}{Phys. Rev. Lett. \textbf{105} , 252301 (2010)}.

\bibitem{Ablyazimov:2017guv}
Ablyazimov, T. and others \textit{et al.} [CBM Collaboration],
``Challenges in QCD matter physics -- the scientific programme of the Compressed Baryonic Matter experiment at FAIR,''
\href{https://link.springer.com/article/10.1140/epja/i2017-12248-y}{Eur. Phys. J. A \textbf{53} , 60 (2017)}.

\bibitem{KEKELIDZE2016846}
V.D. Kekelidze et al.,
``Prospects for the dense baryonic matter research at NICA,''
\href{https://www.sciencedirect.com/science/article/pii/S0375947416300057}{Nucl. Phys. A \textbf{956} , 846-849 (2016)}.

\bibitem{Sako:2014xph}
Sako, H. and others [J-PARC Heavy-Ion Collaboration],
``Studies of high density baryon matter with high intensity heavy-ion beams at J-PARC,''
\href{https://www.sciencedirect.com/science/article/pii/S0375947416300161}{Nucl. Phys. A \textbf{931} , 1158-1162 (2014)}.

%\cite{Baym:2017whm}
\bibitem{Baym:2017whm}
G.~Baym, T.~Hatsuda, T.~Kojo, P.~D.~Powell, Y.~Song and T.~Takatsuka,
``From hadrons to quarks in neutron stars: a review,''
\href{https://iopscience.iop.org/article/10.1088/1361-6633/aaae14}{Rept. Prog. Phys. \textbf{81}  no.5, 056902 (2018)}.

\bibitem{Bzdak2020}
A. Bzdak et al.,
"Mapping the Phases of Quantum Chromodynamics with Beam Energy Scan",
\href{https://doi.org/10.1016/j.physrep.2020.01.005}{Phys. Rept. \textbf{853}, 1-87 (2020)}.

\bibitem{Friman2011}
B. Friman and F. Karsch and K. Redlich and V. Skokov,
"Fluctuations as Probe of the QCD Phase Transition and Freeze-Out in Heavy Ion Collisions at LHC and RHIC",
\href{https://doi.org/10.1140/epjc/s10052-011-1694-2}{Eur. Phys. J. C \textbf{71}, 1694 (2011)}.

\bibitem{Hilger2011}   
T. Hilger and B. Kampfer and S. Leupold,
"Chiral QCD Sum Rules for Open Charm Mesons",
\href{https://doi.org/10.1103/PhysRevC.84.045202}{Phys. Rev. C \textbf{84}, 045202 (2011)}.

\bibitem{Tolos2009}
L. Tolos and D. Cabrera and D. Gamermann and C. Garcia-Recio and R. Molina and J. Nieves and E. Oset and A. Ramos,
"Strange and charm mesons at FAIR",
\href{https://inspirehep.net/files/bbb995438b6f8dd381e485656fe72247}{Acta Phys. Polon. B \textbf{41}, 329-340 (2009)}.

\bibitem{Klingl1999}
F. Klingl and S. Kim and S. H. Lee and P. Morath and W. Weise,
``$J/\psi$ and $\eta_c$ in the nuclear medium: QCD sum rule approach,''
\href{https://doi.org/10.1103/PhysRevLett.82.3396}{Phys. Rev. Lett. \textbf{82}, 3396-3399 (1999)}.

\bibitem{Morita2008}
K. Morita and S. H. Lee,
``Mass shift and width broadening of J/psi in QCD at finite temperature,''
\href{https://doi.org/10.1103/PhysRevLett.100.022301}{Phys. Rev. Lett. \textbf{100}, 022301 (2008)}.




\bibitem{Hilger2009}
T.~Hilger, R.~Thomas and B.~Kampfer,
``QCD sum rules for D and B mesons in nuclear matter,''
\href{https://journals.aps.org/prc/abstract/10.1103/PhysRevC.79.025202}{Phys. Rev. C \textbf{79}, 025202 (2009)}.

\bibitem{Furnstahl:1992ux}
R. J. Furnstahl, D. K. Griegel,  and T. D. Cohen,
`QCD sum rules for nucleons in nuclear matter,''
\href{https://journals.aps.org/prc/abstract/10.1103/PhysRevC.46.1507}{Phys. Rev. C \textbf{46}, 1507-1527 (1992)}.
%\cite{Colangelo:2000dp}
\bibitem{Colangelo:2000dp}
P.~Colangelo and A.~Khodjamirian,
``QCD sum rules, a modern perspective,''
\href{https://arxiv.org/pdf/hep-ph/0010175}{At The Frontier of Particle Physics, 1495-1576 (2001)}

\bibitem{Shifman:1978bx}
Shifman, M. A. and Vainshtein, A. I. and Zakharov, V. I.,
``QCD and Resonance Physics. Theoretical Foundations,''
\href{https://www.sciencedirect.com/science/article/abs/pii/0550321379900221}{Nucl. Phys. B \textbf{147}, 385-447 (1979)}.

\bibitem{Cohen:1991nk}
Cohen, Thomas D. and Furnstahl, R. J. and Griegel, David K.,
``From QCD sum rules to relativistic nuclear physics,''
\href{https://journals.aps.org/prl/abstract/10.1103/PhysRevLett.67.961}{Phys. Rev. Lett.  \textbf{67}, 961 (1991)}.

\bibitem{Bochkarev1986}
A.~I.~Bochkarev and M.~E.~Shaposhnikov,
"The spectrum of hot hadronic matter and finite-temperature QCD sum rules",
\href{https://doi.org/10.1016/0550-3213(86)90209-9}{Nucl. Phys. B \textbf{268}, 220-252 (1986)}.

\bibitem{Ayala2017}
A. Ayala and C. A. Dominguez and M. Loewe,
"Finite Temperature QCD Sum Rules: A Review",
\href{https://onlinelibrary.wiley.com/doi/10.1155/2017/9291623}{Adv. High Energy Phys. \textbf{2017}, 9291623, (2017)}.

\bibitem{Azizi2014} 
K. Azizi and N. Er and H. Sundu,
``More about the B and D mesons in nuclear matter,''
\href{https://doi.org/10.1140/epjc/s10052-014-3021-1}{Eur. Phys. J. C \textbf{74}, 3021 (2014)}.

%\cite{Cobos-Martinez:2025iqg}
\bibitem{Cobos-Martinez:2025iqg}
J.~J.~Cobos-Mart{\'\i}nez, G.~N.~Zeminiani and K.~Tsushima,
``Heavy{\textendash}Heavy and Heavy{\textendash}Light Mesons in Cold Nuclear Matter,''
\href{https://www.mdpi.com/2073-8994/17/5/787}{Symmetry \textbf{17} no.5, 787 (2025)}.

\bibitem{PDG:2024}
S.~Navas \textit{et al.}\ [Particle Data Group],
''Review of Particle Physics,''
\href{https://doi.org/10.1103/PhysRevD.110.030001}{Phys.\ Rev.\ D
\textbf{110}, 030001 (2024)}.

\bibitem{Kumar:2014}
A.~Kumar,
''Heavy Scalar, Vector, and Axial-Vector Mesons in Hot and Dense Nuclear Medium,''
\href{https://doi.org/10.1155/2014/549726}{Adv.\ High Energy Phys.\
\textbf{2014}, 549726 (2014)}.

%\cite{Veliev:2011kq}
\bibitem{Veliev:2011kq}
E.~V.~Veliev, K.~Azizi, H.~Sundu, G.~Kaya and A.~Turkan,
``Thermal QCD Sum Rules Study of Vector Charmonium and Bottomonium States,''
\href{https://link.springer.com/article/10.1140/epja/i2011-11110-8}{Eur. Phys. J. A \textbf{47}, 110 (2011)}.


%\cite{Azizi:2015ona}
\bibitem{Azizi:2015ona}
K.~Azizi and G.~Kaya,
``Modifications on nucleon parameters at finite temperature,''
\href{https://link.springer.com/article/10.1140/epjp/i2015-15172-7}{Eur. Phys. J. Plus \textbf{130}, no.8, 172 (2015)}.

%\cite{Cheng:2007jq}
\bibitem{Cheng:2007jq}
M.~Cheng, N.~H.~Christ, S.~Datta, J.~van der Heide, C.~Jung, F.~Karsch, O.~Kaczmarek, E.~Laermann, R.~D.~Mawhinney and C.~Miao, \textit{et al.}
``The QCD equation of state with almost physical quark masses,''
\href{https://journals.aps.org/prd/abstract/10.1103/PhysRevD.77.014511}{Phys. Rev. D \textbf{77}, 014511 (2008)}.


%\cite{Er:2022cxx}
\bibitem{Er:2022cxx}
N.~Er and K.~Azizi,
``Spectroscopic parameters and electromagnetic form factor of kaon in vacuum and a dense medium,''
\href{https://link.springer.com/article/10.1140/epjc/s10052-022-10333-w}{Eur. Phys. J. C \textbf{82}  no.5, 397 (2022)}.

%\cite{Azizi:2026fnl}
\bibitem{Azizi:2026fnl}
K.~Azizi, N.~Er and J.~Y.~S{\"u}ng{\"u},
`Hot and Dense Medium Effects on the $B_s^*$ and $B^*$ Multiplets,''
\href{https://arxiv.org/pdf/2607.22419}{arXiv:2607.22419 [hep-ph]}.

\bibitem{Dominguez:2007}
C.~A.~Dominguez, M.~Loewe and J.~C.~Rojas,
''Heavy-light quark pseudoscalar and vector mesons at finite temperature,''
\href{https://doi.org/10.1088/1126-6708/2007/08/040}{JHEP \textbf{08},
040 (2007)}.

\end{thebibliography}
\end{document}